\documentclass[aps,10pt,prb,twocolumn,groupedaddress, floatfix, superscriptaddress,showpacs, showkeys,amsfonts]{revtex4-2}
\usepackage[english]{babel}
\usepackage{blindtext}
\usepackage{amsmath,amssymb}
\usepackage{graphicx}
\usepackage[dvipsnames]{xcolor}
\usepackage{hyperref}
\newcommand\numberthis{\addtocounter{equation}{1}\tag{\theequation}}

\definecolor{blue}{HTML}{03045E}
\definecolor{violet}{HTML}{8338EC}
\definecolor{yale_blue}{HTML}{40916C}

\hypersetup{colorlinks=true,linktocpage=true,breaklinks=true,urlcolor=blue,linkcolor=blue,citecolor=blue}

\usepackage[normalem]{ulem}

\usepackage{orcidlink}

\makeatletter
\newcommand{\fmarki}{*}
\newcommand{\fmarkii}{\ensuremath{\dagger}}
\newcommand{\fmarkiii}{\ensuremath{\ddagger}}
\newcommand{\fmarkiv}{\ensuremath{\mathsection}}
\newcommand{\fmarkv}{\ensuremath{\mathparagraph}}
\newcommand{\fmarkvi}{\ensuremath{\|}}
\newcommand{\fmarkvii}{**}
\newcommand{\fmarkviii}{\ensuremath{\dagger\dagger}}
\newcommand{\fmarkix}{\ensuremath{\ddagger\ddagger}}

\def\@fnsymbol#1{{\ifcase#1\or \fmarki\or \fmarkii\or \fmarkiii\or \fmarkiv\or \fmarkv\or \fmarkvi\or \fmarkvii\or \fmarkviii\or \fmarkix \else\@ctrerr\fi}}
\makeatother

\renewcommand{\fmarki}{$\dagger$}
\renewcommand{\fmarkii}{$\ddager$}

\begin{document}

\title{Information scrambling and entanglement dynamics in Floquet Time Crystals}
\author{Himanshu Sahu$^{\orcidlink{0000-0002-9522-6592}}$}
\email{hsahu@pitp.ca}
\affiliation{Perimeter Institute for Theoretical Physics, Waterloo, ON, N2L 2Y5, Canada.}
\affiliation{Department of Physics and Astronomy and Institute for Quantum Computing,University of Waterloo, ON, N2L 3G1, Canada.}
\affiliation{Department of Physics and Department of Instrumentation \& Applied Physics, Indian Institute of Sciences, C.V. Raman Avenue, Bangalore 560012, Karnataka, India.}
\author{Fernando Iemini}
\affiliation{Instituto de Física, Universidade Federal Fluminense, 24210-346 Niterói, Brazil}

\begin{abstract}
We study the dynamics of out-of-time-ordered correlators (OTOCs) and entanglement of entropy as  quantitative measures of information propagation in disordered many-body systems exhibiting Floquet time-crystal (FTC) phases.
We find that OTOC spreads in the FTC with different characteristic timescales due to the existence of a preferred ``quasi-protected'' direction - denoted as $\ell$-bit direction - along which the spins stabilize their period-doubling magnetization for exponentially long times.
While orthogonal to this direction the OTOC thermalizes as an usual MBL time-independent system (at stroboscopic times), along the  $\ell$-bit direction  the system features a more complex structure. The scrambling appears as a combination of an initially frozen dynamics (while in the stable period doubling magnetization time window) and a later logarithmic slow growth (over its decoherence regime) till full thermalization. Interestingly, in the late time regime, since the wavefront propagation of correlations has already settled through the whole chain, scrambling occurs at the same rate regardless of the distance between the spins, thus resulting in an overall envelope-like structure of all OTOCs, independent of their distance, merging into a single growth.
Alongside, the entanglement entropy shows a logarithmic growth over all time, reflecting the slow dynamics up to a thermal volume-law saturation.
\end{abstract}
\maketitle

\section{Introduction}\label{sec:introduction}
Similar to ordinary crystals, time crystals exhibit a high degree of structural order. While ordinary crystals derive their periodicity from the regular repetition of spatial elements, time crystals are an exotic, non-equilibrium state of matter in which the structure arises over time. Both are instances of spontaneous symmetry breaking: the breaking of spatial translation symmetry leads to the formation of crystals, while the breaking of time translation symmetry leads to the formation of time crystals.

Since quantum states in equilibrium have inherently time-independent observables, time-translation symmetry (TTS) arise more naturally in non-equilibrium settings. Indeed, in periodically driven systems, also known as Floquet dynamics, a broad class of models has been shown to break TTS\,\cite{zaletel_colloquium_2023}. In this case, although the system is periodically driven with a Hamiltonian of period T, $\hat H(t) = \hat H(t+T)$, it has observables $O(t)$ (e.g., spin magnetization) that exhibit subharmonic responses, i.e., $O(t) = O(t+nT)$ with $n>1$. Therefore, the system spontaneously breaks the TTS and exhibits a Floquet Time Crystal (FTC) phase.

In general a FTC has an ergodicity breaking mechanism in order to maintain its long-range time ordering, such as induced by disorder in many-body localization (MBL) \cite{PhysRevLett.117.090402,von_keyserlingk_phase_2016,khemani_phase_2016,yao_discrete_2017}, long-range interactions \cite{russomanno_floquet_2017,surace_floquet_2019,yang_dynamical_2021,morita_collective_2006,ojeda_collado_emergent_2021,nurwantoro_discrete_2019,pizzi_higher-order_2021,munoz_arias_floquet_2022,giachetti_fractal_2023}, Stark effect \cite{liu_discrete_2023,lev_discrete_2024}, many-body Scars \cite{maskara_discrete_2021,huang_analytical_2023},
 dissipative dynamics ~\cite{PhysRevLett.121.035301,PhysRevLett.130.180401,Phatthamon2022,PhysRevA.107.L010201,PhysRevLett.130.150401,PhysRevA.107.032219,PhysRevLett.123.260401,Zhu_2019,PhysRevLett.122.015701,Tucker_2018,PhysRevE.97.020202,PhysRevLett.120.040404}, among others \cite{kyprianidis_observation_2021,stasiuk_observation_2023,beatrez_critical_2023,makinen_magnon_2023,euler_metronome_2024}.
In this manuscript we focus on closed MBL periodically driven systems. The many-body localization resulting from strong disorder prevents the quantum system from absorbing the driving energy and heating to infinite temperatures. This new phase of matter has led to several experimental realizations on different platforms, including trapped atomic ions \cite{zhangObservationDiscreteTime2017}, nitrogen-vacancy centers in diamond \cite{choiObservationDiscreteTimecrystalline2017}, superconducting quantum processors \cite{miTimecrystallineEigenstateOrder2022,doi:10.1126/sciadv.abm7652}, spin-based quantum simulators \cite{doi:10.1126/science.abk0603}, and others \cite{PhysRevLett.120.215301,auttiACJosephsonEffect2021}.
Moreover, the use of time crystals for quantum applications has been a subject of growing interest recently, with promissing prospects as quantum sensors~\cite{iemini2023floquet,montenegro_quantum_2023,gribben_quantum_2024,cabot_continuous_2023,yousefjani_discrete_2024,moon2024discretetimecrystalsensing,shukla2024prethermalfloquettimecrystals}, complex network simulators~\cite{Estarellas2020}, quantum engines ~\cite{carollo_nonequilibrium_2020}, among others~\cite{bomantara_simulation_2018,lyu_eternal_2020,bao_schrodinger_2024}.

In non-equilibrium settings, not only the dynamics of local observables is important, but also their correlations and more generally the propagation of information. Interactions can cause the spread of initially localized quantum information into the exponentially many degrees of freedom of the system, a process known as information scrambling~\cite{PRXQuantum.5.010201,lewis-swanDynamicsQuantumInformation2019}. As a consequence, the scrambled information cannot be recovered by local measurements, leading to thermalization~\cite{PhysRevA.43.2046,rigolThermalizationItsMechanism2008,PhysRevE.50.888}. An unambiguous description of this scrambling process remains an open problem. One way to explore it is through out-of-time-ordered correlators ~\cite{swingleUnscramblingPhysicsOutoftimeorder2018a,PRXQuantum.5.010201}, which can help define a quantum analog of the Lyapunov exponent found in classical chaos~\cite{PhysRevLett.118.086801}. This applies to certain systems, typically those in a semi-classical limit or with a large number of local degrees of freedom, which obey a universal upper bound~\cite{maldacenaBoundChaos2016,yasuhirosekinoFastScramblers2008a,lashkariFastScramblingConjecture2013a,Shenker:2013pqa}.

Out-of-time-ordered correlators (OTOCs) allow us to measure the growth of operators. Consider two local operators, $\hat W$ and $\hat V$, in a one-dimensional spin chain. The idea is to study  the spread of $\hat W(t) = e^{i\hat Ht} \hat W e^{-i\hat Ht}$ using another operator $\hat V$, which is a simple spin operator located at some distance $\ell$ from $\hat W$. This can be done by looking at the expectation value of the squared commutator:
\begin{equation}\label{eq.otoc.definition}
C(\ell,t) = \left \langle [\hat W(t),\hat V]^\dagger [\hat W(t),\hat V]\right\rangle/2
\end{equation}
Initially, this value will be zero for well-separated operators, but it will become significantly different from zero once $\hat W(t)$ has spread to the location of $\hat V$. In the special case where $\hat W$ and $\hat V$ are Hermitian and unitary, the squared commutator can be written as $C(\ell,t) = 1 -  \text{Re}[\langle \hat W(t) \hat V \hat W(t) \hat V\rangle]$, with $F(t) = \langle \hat W(t) \hat V \hat W(t) \hat V\rangle $ showing explictly the out-of-time ordering nature of the operators. The growing interest in OTOCs has led to numerous experimental proposals and experiments aimed at measuring them\,\cite{PhysRevA.94.040302,doi:10.1126/science.abg5029,PhysRevLett.129.160602,garttner_measuring_2017,braumuller_probing_2022,landsman_verified_2019}.

For systems with a geometrically local Hamiltonian that do not exhibit localized phase, $C(\ell,t)$ typically exhibits a ballistic growth\,\cite{xu_accessing_2020,PhysRevB.96.020406,Bohrdt_2017,PhysRevB.97.144304}, leading to an emergent linear light cone with a butterfly velocity $v_B$, displaying in this way a rapid growth ahead of the wavefront and a saturation behind it at late times. On the other hand, the growth of $C(\ell,t)$ is severely suppressed in disordered system\,\cite{PhysRevLett.99.167201,PhysRevB.95.060201}. In MBL phases~\cite{PhysRevB.82.174411,RevModPhys.91.021001,annurev:/content/journals/10.1146/annurev-conmatphys-031214-014726,PhysRevX.5.031032,altman_many-body_2018}, characterized by an extensive number of local integrals of motion, the $C(\ell,t)$ exhibits a logarithmic light cone\,\cite{https://doi.org/10.1002/andp.201600332,PhysRevB.95.165136,PhysRevB.95.054201,PhysRevLett.123.165902}.

A different way to track the propagation of information is to examine the buildup of entanglement among the system constituents,  which could be quantified by the entanglement entropy. In a quench system, starting from an unentangled product state, the entanglement entropy initially starts at zero and grows over time. In a thermal system, at late times, the subsystem's density matrix approaches a mixed state, corresponding to a thermal state with a temperature set by the average energy of the initial state. In contrast, in a many-body localized (MBL) system, entanglement entropy grows slowly and saturates at exponentially late times relative to the system size \,\cite{PhysRevLett.109.017202,PhysRevLett.119.110604,PhysRevX.7.021013,RevModPhys.91.021001}.

In this manuscript, we study scrambling and correlation dynamics in a periodically driven disordered spin chain. Specifically, the model consists of interacting nearest-neighbor spins in a disordered chain subject to periodic global spin flips. The system exhibits both an FTC stabilized by MBL as well as a regular ergodic phase for varying its parameters. The scrambling and correlation dynamics over MBL and ergodic phases have been largely discussed in time-independent Hamiltonian systems, it is therefore natural to ask how does the presence of Floquet dynamics, and more specifically of TTS breaking, impacts the phenomenology.

Our study shows that the OTOC growth has different characteristic time scales in the FTC due to the existence of a preferred ``quasi-protected'' direction - so called $\ell$-bit direction. While orthogonal to it the OTOC thermalizes as an usual MBL system, along the localized $\ell$-bit axis the system features a slow scrambling with a combination of frozen dynamics and a later logarithmic slow growth till full thermalization. Interestingly, the late scrambling occurs at the same rate regardless of the distance between the spins, resulting in an overall envelope of all OTOCs independent of their distance. This peculiar behaviour can be interpreted by the fact that at late times the wavefront propagation of correlations has already settled through the whole chain.
The entanglement entropy also features a logarithmically growth over all time, reflecting the slow dynamics till a thermal volume-law saturation.

The paper is structured as follows. We begin with introducing the model in Sec.\,\ref{sec:the model}. In Sec.\,\ref{sec:FTC dynamics}, we present our results for the magnetization dynamics, OTOCs, and entanglement entropy at all time scales. We summarize and conclude with future direction in Sec.\,\ref{sec:discussion}. In Appendix \ref{sec:Exact spin-operator evolution and out-of-time-ordered correlator}  we provide the analytics for the OTOC in absence of perturbative field.

\begin{figure*}
\centering
\includegraphics[width = 0.322\linewidth]{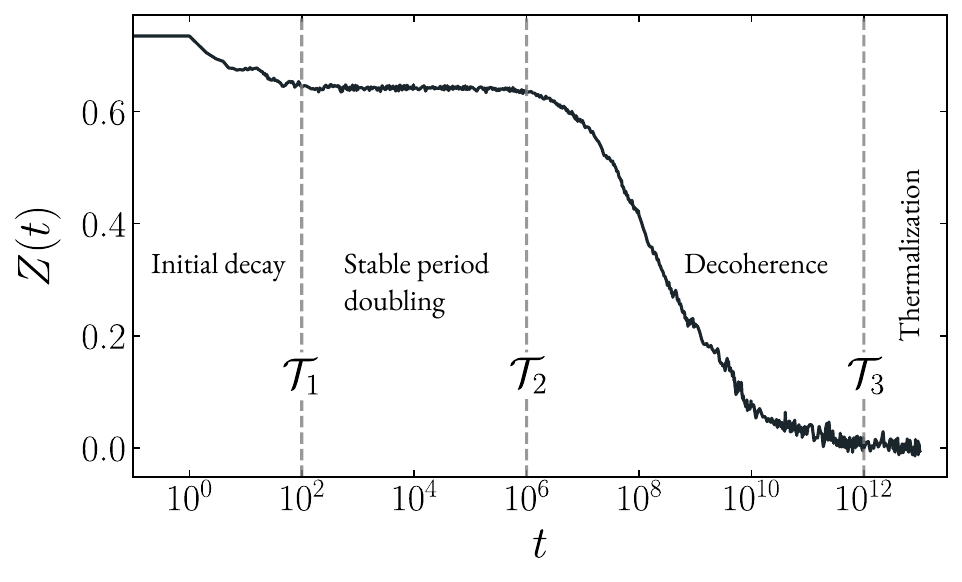}
\includegraphics[width = 0.32\linewidth]{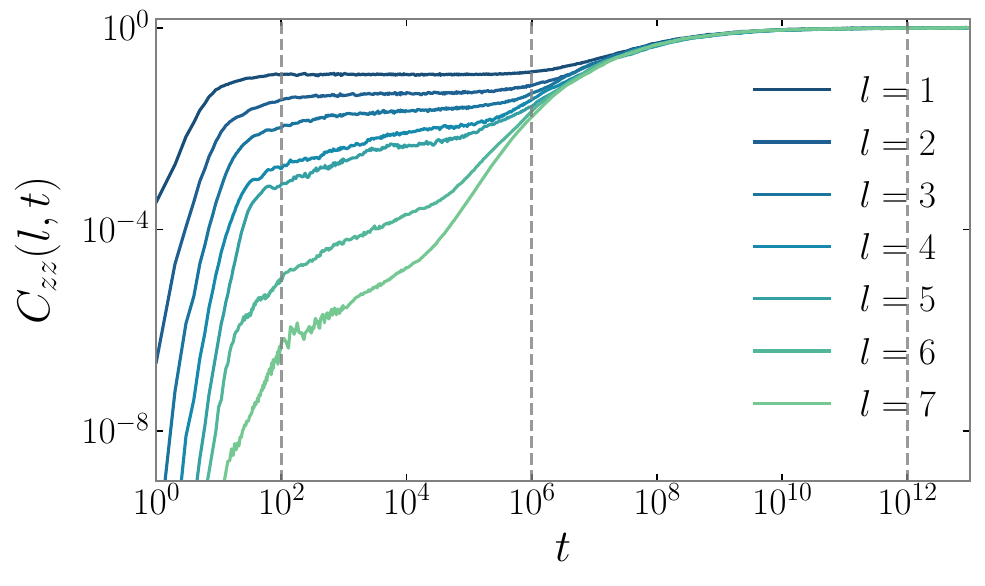}
\includegraphics[width = 0.33\linewidth]{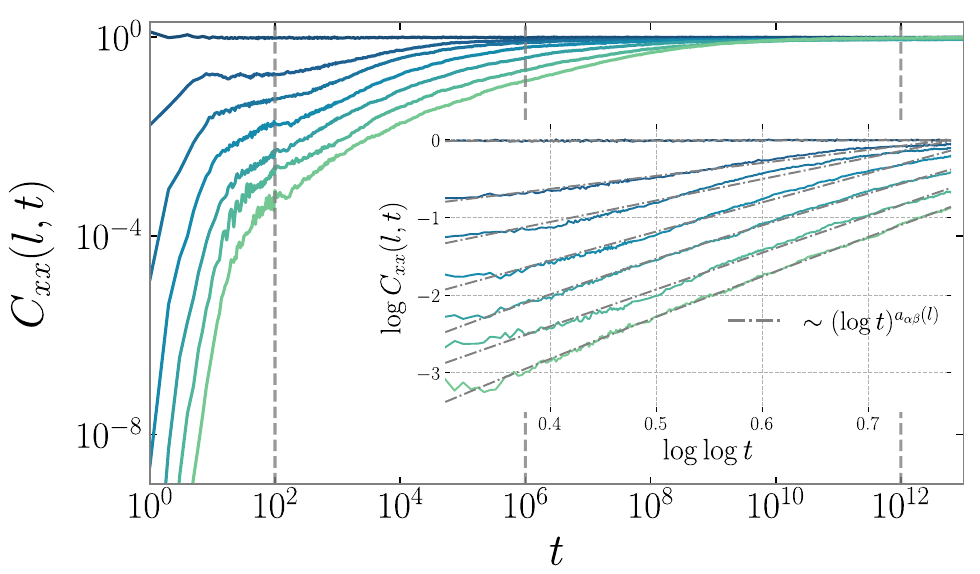}
\caption{\textbf{(Left)} The dynamics of magnetization $Z(t)$ in the FTC phase, and its different characteristic time scales $\mathcal{T}_i$ (vertical dotted lines). The magnetization displays a period doubling plateau in the fully stable regime of the time crystal, which later slowly decoheres and thermalises at late times.  We show the corresponding dynamics of OTOCs along the \textbf{(center)} $z$-direction ($C_{zz}(l,t)$) and \textbf{(right)} $x$-direction ($C_{xx}(l,t)$), for varying spin distances $l$. The inset-panel shows the dynamics within the regime  $\mathcal{T}_1<t<\mathcal{T}_2$ in a semi-log scale in order to highlight the logarithmic growth of the OTOC, where the dotted curves are the fitted plots using the function $(
\log t)^{a_{\alpha \beta}(l)}$. In all results we used $L = 8$ spins and  coupling parameters $J = 1$, $h^{x} = 0.2$, and $h^{z} = 1$. The operator distances $l$ are always taken from the bulk of the spin chain to the edges (most distant spins).}
\label{fig:OTOC}
\end{figure*}

\section{The Model}\label{sec:the model}

We consider a one-dimensional spin-$1/2$ chain with a Floquet unitary of the form
\begin{equation}\label{eq.UF} 
\hat U_F  = \exp(-i \hat H_0 T) \, \hat U_K, \quad
\hat U_K \equiv \exp\left( i\phi  \sum_{i=1}^L \hat{\sigma}_i^x/2\right),
\end{equation}
where $L$ is the number of spins in the chain, $T$ the period of the Floquet dynamics, $\phi$ the kicking phase and the Hamiltonian is defined as,
\begin{equation}\label{eq:HMBL}
\hat H_0 = \sum_i \left(J_i \hat{\sigma}_i^z\hat{\sigma}^z_{i+1} + h_i^z \hat{\sigma}^z_i + h^x_i \hat{\sigma}_i^x\right)
\end{equation}
with $\hat{\sigma}_i^\alpha$ the Pauli operator for the $i$'th spin along the $\alpha=x,y,z$ direction. The couplings
$J_i, h_i^z$, and $h_i^x$ are uniformly chosen from $J_i \in [J/2,3J/2]$ , $h_i^z \in [0,h^z]$, $h_i^x \in [0,h^x]$. For strong enough disordered interaction ($J_i$)  and in case of no kicking ($\phi=0$), the system evolves to an MBL state. In case of periodic kickings close to $\phi \backsimeq \pi $ the dynamics  stabilizes a FTC phase~\cite{PhysRevB.107.115132}. The FTC displays a robust period doubling magnetization dynamics, which was observed in various experiments\,\cite{choiObservationDiscreteTimecrystalline2017,zhangObservationDiscreteTime2017,doi:10.1126/science.abk0603,doi:10.1126/sciadv.abm7652,miTimecrystallineEigenstateOrder2022,xu2021realizingdiscretetimecrystal}.

\section{FTC dynamics}\label{sec:FTC dynamics}
In this section, we show our results for the dynamics of magnetization, OTOCs, and entanglement entropy in the FTC phase. Their behaviour follow different growth/decay rates at corresponding characteristic time scales, which depend on the coupling parameters and system size, as we discuss in detail. In all our simulations we consider initial product states $\left[\cos \theta \lvert \uparrow \rangle + \sin \theta \lvert \downarrow \rangle \right]^{\otimes L}$, with $\theta = \pi/8$, open boundary conditions and average our results over $10^3$ disorder simulations. However, our results are independent of the choice of initial state.

\subsection{Magnetization dynamics}\label{subsec:magnetization dynamics}

The period doubling dynamics can be computed analytically in the simpler case of no transverse field $h^x = 0$ and a kicking phase $\phi = \pi$, where $\hat U_K =  \prod_i i\hat{\sigma}^x_i$ has the effect of perfectly flipping all the spins.
In this case the Hamiltonian $\hat H_\text{0}$ is diagonal in eigenbasis ($\lvert \uparrow\rangle $ and $\lvert \downarrow\rangle $) of individual Pauli $\hat{\sigma}^{z}_i$ operator. Each eigenstate of the Hamiltonian $H_\text{0}$ can be written as $|s_1s_2 \ldots \rangle \equiv |\{s_i\}\rangle$ with $s_i =\pm 1$, so that $\hat{\sigma}^z_k|\{s_i\}\rangle = s_k|\{s_i\}\rangle$. The action of the Floquet unitary $\hat U_F$ on a random initial product state is to flip each spin, i.e. $\hat U_F |\{s_i\}\rangle = e^{-iE(\{s_i\})T}\lvert \uparrow \downarrow\downarrow\uparrow\downarrow \cdots \rangle$ where $E(\{s_i\})$ is energy eigenvalue of state $|\{s_i\}$ associated with Hamiltonian $\hat H_\text{0}$. The magnetization $\langle \hat{\sigma}^z_i(t)\rangle $ oscillates with period $2T$, while the magnetization
\begin{equation}
Z(t) = (-1)^t \langle \hat{\sigma}^z_i(t)\rangle \text{sgn} \left(\langle \hat{\sigma}_i^z(0)\rangle \right)
\end{equation}
takes a constant value. Most importantly, for large system sizes $L$ and disordered interactions $J_i$, the period doubling dynamics is stable under perturbations of the Floquet operator such as a non-zero transverse field $h^{x}$ or imperfect spin flip $\phi = \pi + \epsilon$~\cite{PhysRevLett.117.090402}. Therefore, the system breaks the time-translation symmetry stabilizing a FTC phase.

In the general case, the magnetization dynamics in FTCs for finite-system sizes can be described by different dynamical regimes. We illustrate them in Fig.\eqref{fig:OTOC}-(left panel).
The magnetization exhibits in increasing order of time:

$\bullet \ \, 0 <t<\mathcal{T}_1$: a first initial spin relaxation leading to a decay in the magnetization. This first dynamical regime is general and roughly independent on the spin microscopic details, i.e., to an MBL/ergodic/FTC phase;

$\bullet \ \, \mathcal{T}_1 <t<\mathcal{T}_2$: a stable period doubling dynamics till time $\mathcal{T}_2$, with all the spins now in a coherent dynamics due to kicking and MBL interactions. Due to MBL, the time scale for period doubling dynamics increases exponentially with the system size, $\mathcal{T}_2 \sim e^{\beta L}, \beta >0$, therefore persisting indefinitely in the macroscopic limit, $L \rightarrow \infty$.

$\bullet \ \, \mathcal{T}_2 <t<\mathcal{T}_3$: for longer times, the spins now start to decohere among each other and the magnetization $Z(t)$ slowly decays.
We find a logarithmic decay leading to a characteristic time scale exponentially long with the system size,  $\mathcal{T}_3 =e^{\beta' L}$. In particular, for the chosen parameters, a time exponent even larger than the period doubling regime, with $\beta' \approx 0.95 > \beta \approx 0.43$.

$\bullet\  \, \mathcal{T}_3 <t$: in the late time, the system has now thermalized, with a null net magnetization, $Z(t) \rightarrow 0$, and trivial dynamics.

\subsection{ Out-of-time-ordered correlator}\label{subsec:otoc}
We now turn our attention to the behavior of OTOCs (Eq.\,\eqref{eq.otoc.definition}).
We consider the scrambling of local spin operators: $\hat W(0), \hat V \in \{ \hat \sigma_i^\alpha\}$. In order to simplify the notation, we define
\begin{equation}\label{eq:OTOC}
C_{\alpha \beta}(\ell,t) \equiv \frac{1}{2} \left\langle \left| [\hat \sigma_\ell^\alpha(t), \hat \sigma^\beta_1] \right|^2 \right\rangle
\end{equation}
where $\ell = 1,...,L$, and  $\alpha,\beta \in \{x,y,z\}$. We focus on $C_{xx}(\ell,t), C_{zz}(\ell,t)$ as these are representative of the main behaviors of the OTOC in the system.

In the case of a perfect spin-flip ($\phi = \pi$) with no transverse field ($h^x=0$), the local operators along the different directions remain local under the Floquet unitary. Therefore the OTOC for distant spins is constant over time (see Appendix~\ref{sec:Exact spin-operator evolution and out-of-time-ordered correlator}). This is a fine-tuning point of the model (limit of vanishing correlation length), for which there is no scrambling, and it is not representative of the general behavior.

In the general FTC case, the scrambling dynamics  display different characteristic time scales $\mathcal{T}_i$, whose main features are strongly influenced by the spin directions - see Fig.\eqref{fig:OTOC} - as we discuss below.

$\bullet\, \, 0 <t<\mathcal{T}_1$: the initial OTOC growth arises due to the spins initial relaxation - similar to magnetization dynamics - and it is independent on the underlying non-equilibirum phase;

$ \bullet\, \, \mathcal{T}_1 <t<\mathcal{T}_2$: despite in a stable period doubling magnetization, the OTOC has a slow growth in this regime.
Its main behavior (apart from small corrections, discussed  later) is similar to an \textit{MBL-like dynamics}. I.e., its growth resembles those of a quench dynamics in a time-independent MBL spin system\,\cite{PhysRevB.99.184202,FAN2017707}. After the initial transient time $\mathcal{T}_1$, the OTOC shows a logarithmically scrambling of information, with
\begin{equation}
C_{\alpha \beta}(\ell,t) \sim A_{\alpha \beta}(\ell) \log(t)^{a_{\alpha \beta}(\ell)},
\end{equation}
where $A_{\alpha \beta}(\ell) \sim e^{-v_{\alpha \beta} \ell}$ - see inset  of Fig.\eqref{fig:OTOC}-(right panel). The exponents $a_{\alpha \beta}(\ell), v_{\alpha \beta}(\ell)$ are non-universal, depending on the distance and system interactions. A few observations are in order. While both $C_{xx}(\ell,t)$ and $C_{zz}(\ell,t)$ show a slow growth, one clearly notice that the scrambling along the $x$-direction is much larger (by orders of magnitude) than the one along the $z$-direction. Indeed, while  closer spins ($\ell \lesssim3$) $C_{xx}(\ell,t)$ may already get fully scrambled in this time window, $C_{zz}(\ell,t)$ has generally rather small values, remaining close to \textit{rough plateau’s} depending on their spin distances. These small values arise due to the (period doubled) conservation of the $z$-magnetization,
which causes only a small part of $\hat \sigma_i^z$ to scramble, leaving the most of it protected during the dynamics. On the other hand, the magnetization along the $x$-direction has no protection or stability in the FTC, i.e., it is vanishingly small even within the period doubling time regime, with expectation values resembling those of a high temperature state, therefore leading to a larger (in magnitude) scrambled $\hat \sigma_i^{x}$ spin operator.

Alternatively, one could also understand this regime by recalling an $\ell$-bit phenomenological model for \textit{MBL-like dynamics}\,\cite{PhysRevB.90.174202,ROS2015420,PhysRevB.91.161109,PhysRevLett.111.127201,imbrie_many-body_2016}. The picture \textit{assumes} the existence of local integrals of motion along a preferred direction, denoted as $\ell$-bit operators, $\hat \tau_i^z$. In the limit of infinite localization, one expects the local spin operators to be roughly aligned to the integral of motion, as e.g. in the simplest FTC case with perfect spin-flip ($\phi=\pi$) and no transverse field ($h^x=0$), where we easily identify $\hat \sigma_i^z = \hat \tau_i^z$.  On the general case though (not infinite localization), the spin operators $\hat \sigma_i^\alpha$ have an overlap both into more distant $\ell$-bit operators ($\hat \tau_{j \neq i}^\alpha$) as well as onto different orthogonal $\hat \tau_j^x$ and $\hat \tau_j^y$ directions, which are \textit{not} integrals of motion. Specifically,
\begin{equation}\label{eq.lbit.decomp}
\hat \sigma_i^\alpha = \sum_{j,\beta} {c}_{j,\beta}^{[i,\alpha]} \hat \tau_{j}^{\beta} + \sum_{j k} {c}_{j k,\beta \gamma}^{[i,\alpha]} \hat \tau_{j}^{\beta} \hat \tau_{k}^{\gamma} + ...
\end{equation}
with the coefficients ${c}_{...}^{[i,\alpha]}$  exponentially localized around the $i$’th position, corresponding to the overlap into the new locally deformed $\{\hat \tau_i^\alpha\}$ basis. The last ``$\cdots$'' term in the equation represents possible higher order products of $\ell$-bit operators.
Therefore, as long as the spin operator has an overlap into an orthogonal direction - $\hat \tau_j^{x,y}$ -  part of it scrambles, and this explains the difference between $C_{xx}(\ell,t)$ and $C_{zz}(\ell,t)$ behaviors. Both $\hat \sigma_i^z$ and $\hat \sigma_i^x$ spin operators have, in general, an overlap onto the
the orthogonal $\ell$-bit operators. However, while spin $\hat \sigma_i^z$ has a vanishing small overlap, which decreases with the localization strength, $\hat \sigma_i^x$ has almost its totality onto it. Therefore, while only the small overlapping fraction of $\hat \sigma_i^z $ is scrambled, $\hat \sigma_i^x $ is scrambled on its majority.

\begin{figure}
\centering
\includegraphics[width = 0.98\linewidth]{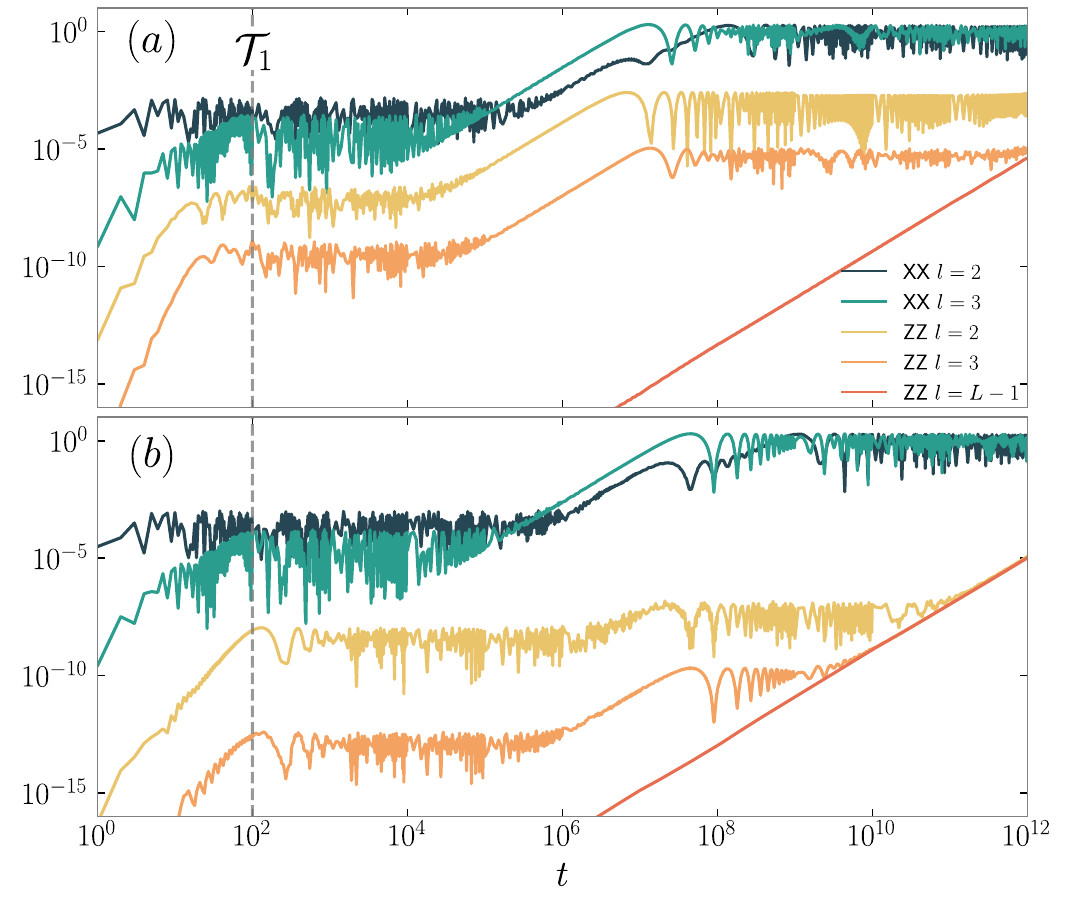}
\caption{The dynamics of OTOCs $C_{\alpha \beta}(l,t)$ for single realizations. We set the system deep in the FTC phase (with parameters $J=h^z = 1$, $h^x = 0.025$ and $L = 8$ spins) highlighting the OTOC in the (long) period doubling regime, $t<\mathcal{T}_2$. We show in panels (a) and (b)  two typical dynamics for different single disorder realizations. While different OTOC directions show a qualitative similar behavior (apart from their orders of magnitude), the two instances of disorder realizations display different plateau time windows.}
\label{fig:OTOC Deep MBL}
\end{figure}

In order to corroborate this picture, we show in Fig.\eqref{fig:OTOC Deep MBL} the OTOC dynamics in the period doubling magnetization time ($t<\mathcal{T}_2$). We set the system in the strong localization regime in order to enlarge $\mathcal{T}_2$ highlighting its main features, and show the OTOC for individual disorder realizations along $x$ and $z$-directions. We observe that, after the initial relaxation time ($\mathcal{T}_1$), both $C_{xx}(\ell,t)$ and $C_{zz}(\ell,t)$ show qualitatively similar behavior, with the main difference on their order of magnitudes. This is explained due to the fact that both spin operators $\hat \sigma_i^x$ and $\hat \sigma_i^z$ can share the same orthogonal $\ell$-bit terms on their decomposition (Eq.\eqref{eq.lbit.decomp}), however, with different weight coefficients, which leads to the different orders of magnitude in the scrambled operator.
Interesting to notice, at individual disorder realizations, the OTOC dynamics features plateau's interspaced by quadratic growths ($\sim t^2$). Since the plateau widths are disorder dependent, once one averages over the disorder realizations, this behavior is hidden, taking space to the logarithmic growth~\cite{PhysRevB.99.184202}.

$\bullet\,  \, \mathcal{T}_2 <t<\mathcal{T}_3$: for longer times the spins decohere among each other and the FTC magnetization is slowly melted. While correlators along orthogonal directions to the $\ell$-bit protected direction (i.e., $C_{xx}(\ell,t)$) keeps its logarithmic growth till saturation, with no significant differences, the scrambling along the prefered $\tau_i^z$ direction has a rather peculiar behavior, as captured by $C_{zz}(\ell,t)$. The correlator $C_{zz}(\ell,t)$ at these late times ($t \lesssim \mathcal{T}_2$) has roughly saturated to a value much lower than its upper bound, with the saturated value related to the protected overlap to $\hat \tau_i^z$ operators. However, now due to decoherence, when the FTC starts to melt (no protected $\ell$-bit direction anymore), the correlator starts to grow again similar to thermal systems. An important observation is that, since the correlation wavefront (``lightcone'') has already passed along all the chain, the thermal growth is now independent of the distance $\ell$; therefore, the correlator $C_{zz}(\ell,t)$ have all the same growth and merge into a single one forming an envelope-like structure\,\cite{xu_accessing_2020}. Precisely, in this time window the correlator $C_{zz}(\ell,t) \sim e^{-b_\ell \ell} + c \log(t)^d$, with $c,d$ independent on the distance $\ell$, where the first term stands for the saturation value achieved in the period doubling time window, and the second term arises from the onset of thermalization in the decoherence process.

$\bullet\  \, \mathcal{T}_3 <t$: in the late time the system now is full thermal, and the OTOCs along all directions reach their maximun value.

\subsection{Entanglement dynamics}\label{subsec:entanglement dynamics}
We compare the scrambling of information to the dynamics of entanglement in the spins. In order to quantify the correlations we consider the entanglement entropy in an equal bipartition of the system. Specifically, given the density matrix $\hat \rho = |\Psi\rangle \langle \Psi |$, the total system is divided into left (A) and right (B) subsets, each with half of the total spins in the chain and corresponding reduced states $\hat \rho_{A(B)} = \text{Tr}_{B(A)}( |\Psi\rangle \langle \Psi |)$. The entanglement between the bipartition is defined by the von Neumann entropy of any of the reduced state,
\begin{equation}
S =  -\frac{\text{Tr}\left[\hat \rho_{A} \ln(\hat \rho_A)\right]}{L_A}
\end{equation}
where $L_A=L/2$ is size of subset $A$. In a thermal system, the entanglement entropy of an initial low entangled state typically increases linearly over time until it reaches the value associated with the system's energy density, where the entanglement entropy obeys the thermodynamic volume law. Conversely, in a localized system, one may expect that since the information remains confined near the subsystem, the entanglement entropy should saturate at a constant value proportional to the localization length rather than the subsystem size. In fact, in AL system, the entanglement entropy saturates to a constant independent of the system size. In generic MBL systems, however, it has been observed\,\cite{PhysRevLett.109.017202,PhysRevLett.119.110604,PhysRevX.7.021013,RevModPhys.91.021001} that the entanglement entropy grows slowly, following a $\log t$ dependence in time, and saturates to a volume law at exponentially long times.

\begin{figure}
\centering
\includegraphics[width = 0.95\linewidth]{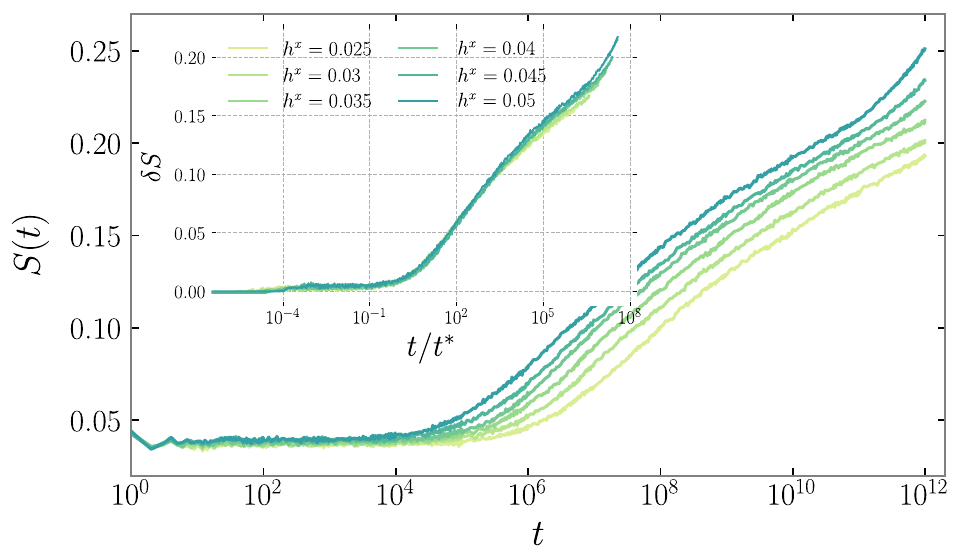}
\caption{ Dynamics of entanglement entropy $S(t)$ deep in the FTC phase ($h^x\approx 0$), for varying $h^{x}$ fields with a fixed system size $L = 8$, and couplings $J= h^z= 1$. The inset shows the subtracted entanglement entropy $\delta S = S_{h^x}(t) - S_{h^x=0}(t)$ with a rescaled time $t/t^*$.}
\label{fig:Entropy_Deep_MBL_Comp}
\end{figure}

\begin{figure}
\centering
\includegraphics[width = 0.95\linewidth]{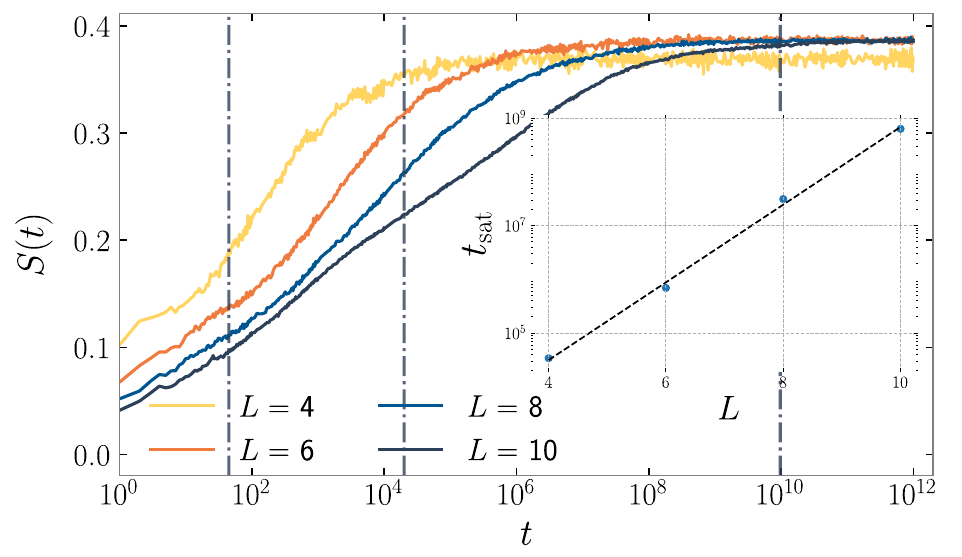}
\caption{Late-time dynamics of entanglement entropy in the FTC phase, for varying system sizes $L$. The inset panel shows the exponential scaling of saturation time $t_\text{sat}$ with the number of spins. The vertical dotted lines represent the characteristic time scales $\mathcal{T}_{1,2,3}$, respectively, for $L=10$ spins. We set system parameters as $J = 1 = h^z$ and $h^x = 0.5$.
}
\label{fig:late-time-entropy}
\end{figure}

In Fig.~\ref{fig:Entropy_Deep_MBL_Comp}, we show the entanglement entropy calculated in deep FTC phase $h^x \approx 0$. In this regime, the entanglement growth obeys similar behavior as observed in a MBL system\,\cite{PhysRevLett.109.017202}. The entanglement entropy remains constant and equal to the no transverse field case, $S(h^x = 0)$, up to a growth time $t^*$. The growth time is exponentially suppressed with the perturbative field $h^x$, specifically, as $\log t^* \propto (h^x)^{-p}$ where $0< p\leq 1$. Therefore, when we renormalize the time axis with respect to growth time $t^*$, all curves roughly merge onto a single one (see inset Fig.\,\ref{fig:Entropy_Deep_MBL_Comp}). A similar universality is seen in different disordered system with MBL phase\,\cite{PhysRevLett.109.017202}. For time $t>t^*$, the entanglement shows a slow logarithmic growth, typical in the MBL. To probe the late-time entanglement dynamics, we consider larger perturbative $h^{x}$ fields. As previously discussed, the magnetization dynamics tends to a full thermalization at sufficiently late times. We find that the logarithmic growth extends till saturation time  (See Fig.~\ref{fig:late-time-entropy}). The saturation time $t_\text{sat}$ grows exponentially with system-size \textit{i.e.} $\log t_\text{sat}\sim L$, whose value obeys a volume-law.

\smallskip
\section{Discussion}\label{sec:discussion}

In summary, we have investigated the dynamics of OTOCs and entanglement entropy in a disordered spin chain exhibiting Floquet time-crystal (FTC) phases. Our findings reveal distinct characteristics of information propagation within this system, governed by the presence of a quasi-protected $\ell$-bit direction. OTOCs exhibit different timescales for growth in the FTC phase, reflecting the interplay between frozen dynamics and slow logarithmic scrambling along the $\ell$-bit axis. Orthogonal to this direction, the dynamics resemble those of a conventional MBL system, with thermalization occurring at stroboscopic times. However, along the $\ell$-bit direction, a unique behavior emerges, where the system retains a stable period-doubling magnetization for exponentially long times before entering a decoherence regime characterized by logarithmic growth. At late times, the wavefront propagation of correlations completes, resulting in a uniform scrambling rate across all distances, and OTOCs converge into a single growth profile, independent of separation. Additionally, the entanglement entropy displays a persistent logarithmic increase, consistent with the slow dynamics of the system, eventually reaching a thermal volume-law saturation. 

The experimental measurement of OTOCs is feasible in Floquet time crystal due to common platforms where OTOCs have been measured and where FTC have been realized. As an example, OTOCs were experimentally measured in superconducting quantum processors using a scheme that utilizes an interferometric protocol mapping correlators $C(\ell,t)$ onto the projection of an ancilla qubit\,\cite{doi:10.1126/science.abg5029,PhysRevA.94.040302}. On the other hand, the FTC was also experimentally implemented on an array of superconducting qubits using tunable controlled-phase gates where, in fact, the same interferometric protocol was used to quantify the impact of external decoherence\,\cite{miTimecrystallineEigenstateOrder2022}. The observation of OTOCs in FTCs are thus promissing in such platforms; other possibilities include but are not limited to trapped atomic ions, spin-based systems, optical-lattice, neutral atoms, among others.

Future theoretical investigations could explore how scrambling manifests in other types of FTCs, such as those stabilized by Stark many-body localization \cite{PhysRevLett.130.120403} or quantum many-body scars \cite{PhysRevLett.129.140602,PhysRevLett.127.090602}. While this study primarily utilizes OTOCs as a probe for scrambling, alternative measures, such as Krylov complexity \cite{PhysRevX.9.041017}, could provide additional insights. Krylov complexity, which quantifies the delocalization of a local operator under Heisenberg evolution, has been studied in previous works \cite{Sahu:2024urf,10.21468/SciPostPhys.13.2.037,PhysRevA.110.052405} to characterize localized phases in both continuous and discrete-time systems. Extending this approach to FTCs could reveal whether and how the behavior of Krylov complexity differs from that observed in MBL systems.

\begin{acknowledgments}
Research at Perimeter Institute is supported in part by the Government of Canada through the Department of Innovation, Science and Economic Development and by the Province of Ontario through the Ministry of Colleges and Universities. F.I. acknowledges financial support from the Brazilian funding agencies CAPES, CNPQ, and FAPERJ (Grants No. 308205/2019-7, No. E-26/211.318/2019, No. 151064/2022-9, and No. E-26/201.365/2022) and by the Serrapilheira Institute (Grant No. Serra 2211-42166).
\end{acknowledgments}

\appendix

\section{Exact OTOC evolution}\label{sec:Exact spin-operator evolution and out-of-time-ordered correlator}

In case of perfect FTC phase corresponding to parameter $h^{x} = 0$ and $\phi = \pi/2$ (we will further take $T=1$ throughout), the evolution of spin-operators and therefore, OTOC can be calculated exactly. The evolution of spin-operator $\hat{\sigma}_l^z$ ($l>0$) defined as
\begin{equation}
\hat{\sigma}_l^z(1) = (\hat{U}_F\hat{U}_K)^\dagger \hat{\sigma}_l^z (\hat{U}_F\hat{U}_K) = \hat{U}_K^\dagger \hat{U}_F^\dagger \hat{\sigma}_l^z \hat{U}_F \hat{U}_K\,.
\end{equation}
Using $[\hat{U}_F,\hat{\sigma}_l^z] = 0$ and $\hat{U}_K^\dagger \hat{\sigma}_l^z \hat{U}_K = -\hat{\sigma}^z_l$ --- we have $\hat{\sigma}_l^z(t) = (-1)^t \hat{\sigma}_l^z$. Therefore, the $zz$ OTOC $F_{zz}(l,t) = \langle \hat{\sigma}^z_l(t) \hat{\sigma}_0^z \hat{\sigma}^z_l(t) \hat{\sigma}^z_0 \rangle = 1$ and  $C_{zz}(t) = 0$ showing no scrambling in $z$-direction. The evolution of spin-operators $\hat{\sigma}^x_l$ given by
\begin{widetext}
\begin{align*}
\hat{\sigma}_l^x(1) &= \hat{U}^\dagger_K \hat{U}^\dagger_F \hat{\sigma}_l^x\hat{U}_F \hat{U}_K = \sum_{ij} \exp\left[i (E( s_j )-E(s_i)\right] \langle \{s_j\} |\hat{\sigma}_l^x|\{s_i\}\rangle \cdot  \hat{U}^\dagger_K|\{s_j\}\rangle \langle \{s_i\}|  \hat{U}_K
\end{align*}
If we represent the state $|\{s_i\}\rangle = |i_0 i_1\ldots i_l\ldots \rangle$ so that
\begin{equation}
\langle \{s_j\} |\hat{\sigma}^x_l |\{s_i\}\rangle = \delta_{i_0j_0}\delta_{i_1j_1}\ldots \delta_{\bar{i}_lj_l}\ldots
\end{equation}
where $i_k = \pm 1$ and $\overline{\pm 1} = \mp 1$. With this, we can write
\begin{equation}
    \begin{split}
        \hat{\sigma}^x_l(1) &= \sum_i \exp[i(E(i_0i_1\ldots \bar{i}_l\ldots )  - E(i_0i_1\ldots i_l\ldots ))] |\bar{i}_0\bar{i}_1\ldots i_l\ldots \rangle\langle \bar{i}_0\bar{i}_1\ldots \bar{i}_l\ldots | \\
        &= \sum_i e^{i \epsilon_i } |\bar{i}_0\bar{i}_1\ldots i_l\ldots \rangle\langle \bar{i}_0\bar{i}_1\ldots \bar{i}_l\ldots |
    \end{split}
\end{equation}
where $\epsilon_i = E(i_0i_1\ldots \bar{i}_l\ldots )  - E(i_0i_1\ldots i_l\ldots)=\Delta  i_l  \left(J_l i_{l+1} + J_{l-1}i_{l-1} + h^z_l\right) $ with $\Delta i_l = \bar{i}_l - i_l = \mp 2$. Following the same steps,

\begin{equation}
    \begin{split}
        \hat{\sigma}^x_l(2) &= \sum_i e^{i \epsilon_i } \exp[i(E(\bar{i}_0\bar{i}_1\ldots i_l\ldots )  - E(\bar{i}_0\bar{i}_1\ldots \bar{i}_l\ldots ))] |i_0 i_1 \ldots \bar{i}_l \ldots \rangle \langle i_0 i_1 \ldots i_l \ldots | \\
        &= \sum_i e^{i (\epsilon_i- \bar{\epsilon}_i) }  |i_0 i_1 \ldots \bar{i}_l \ldots \rangle \langle i_0 i_1 \ldots i_l \ldots |
    \end{split}
\end{equation}
where $\bar{\epsilon_i} = - E(\bar{i}_0\bar{i}_1\ldots i_l\ldots )  + E(\bar{i}_0\bar{i}_1\ldots \bar{i}_l\ldots ) = \Delta i_l (J_l \bar{i}_{l+1} + J_{l-1}\bar{i}_{l-1} + h^z_l)$. In general,
\begin{equation}
\hat{\sigma}^x_l (t)  =\sum_i e^{ \pm i (n\epsilon_i- m\bar{\epsilon}_i) }  |i_0 i_1 \ldots \bar{i}_l \ldots \rangle \langle i_0 i_1 \ldots i_l \ldots |
\end{equation}
where $n = \lfloor (t+1)/2\rfloor$, $m= \lfloor t/2\rfloor$ and $\pm$ taken for even and odd time-steps, respectively. We can compute the $xx$ OTOC correlator as ($l >1$)

\begin{figure*}
\centering
\includegraphics[width = 0.322\linewidth]{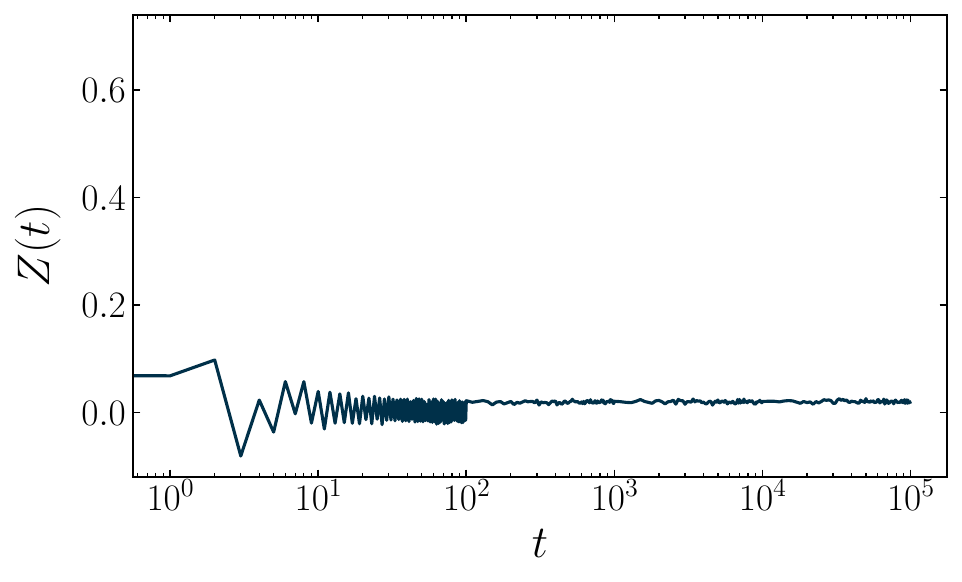}
\includegraphics[width = 0.32\linewidth]{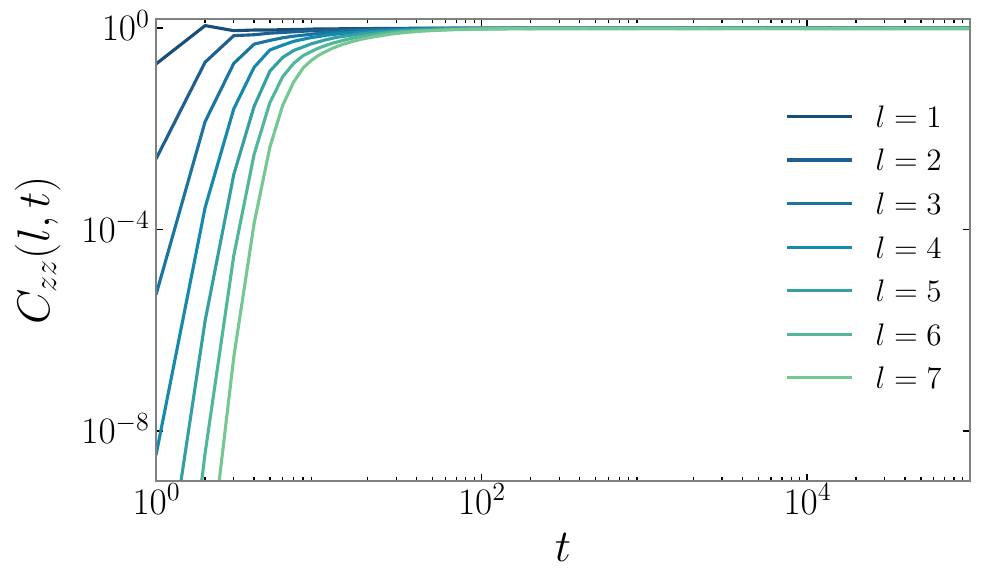}
\includegraphics[width = 0.33\linewidth]{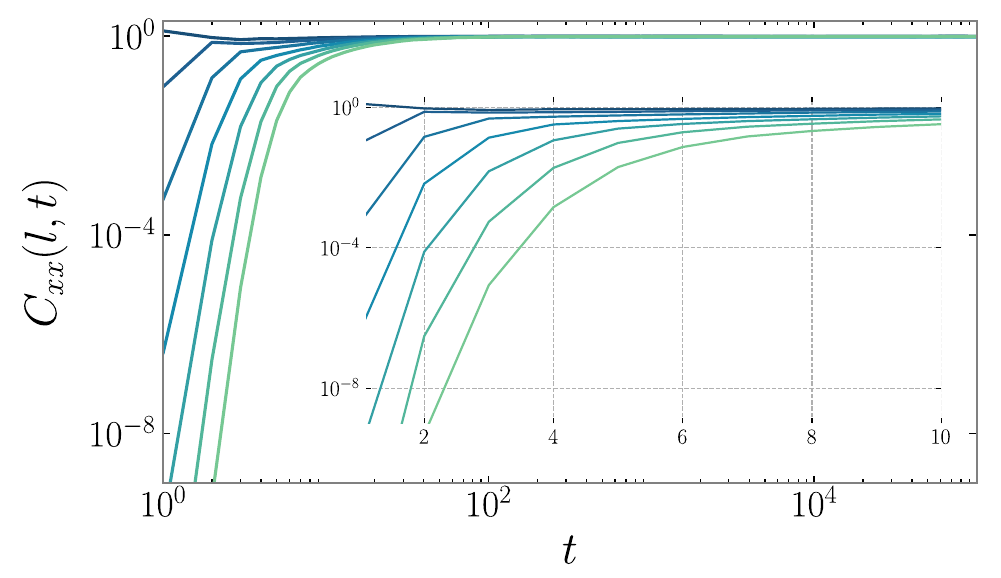}
\caption{The dynamics associated with melted FTC (ergodic phase) for system-size $L = 8$ and parameter values $J=1 =h^z$, $h^x = 0.5$ and the kicking phase $\phi = \pi/2 + 0.4$. \textbf{(Left)} The magnetization $Z(t)$, \textbf{(center)} the OTOCs along the $z$-direction ($C_{zz}(l,t)$) and \textbf{(right)} $x$-direction ($C_{xx}(l,t)$), for varying spin distances $l$. The inset shows the early time growth.}
\label{fig:FTC Melting}
\end{figure*}

\begin{align*}
\hat{\sigma}^x_l(t) \hat{\sigma}_0^x \hat{\sigma}^x_l(t) \hat{\sigma}^x_0 &= \sum_{ij} e^{ \pm i (n\epsilon_i- m\bar{\epsilon}_i) } e^{ \pm i (n\epsilon_j- m\bar{\epsilon}_j) }|i_0 i_1 \ldots \bar{i}_l \ldots \rangle  \langle i_0 i_1 \ldots i_l \ldots | \hat{\sigma}^x_0   |j_0 j_1 \ldots \bar{j}_l \ldots \rangle   \langle j_0 j_1 \ldots j_l \ldots | \hat{\sigma}^x_0 \\
&= \sum_{ij} e^{ \pm i (n\epsilon_i- m\bar{\epsilon}_i) } e^{ \pm i (n\epsilon_j- m\bar{\epsilon}_j) }|i_0 i_1 \ldots \bar{i}_l \ldots \rangle  \langle i_0 i_1 \ldots i_l \ldots  | \bar{j}_0 j_1 \ldots \bar{j}_l \ldots \rangle   \langle \bar{j}_0 j_1 \ldots j_l \ldots |  \\
&= \sum_{j} | \bar{j}_0 j_1 \ldots j_l \ldots \rangle  \langle \bar{j}_0 j_1 \ldots j_l \ldots | = I
\end{align*}
it follows that $F_{xx}(t) = \langle \hat{\sigma}_l^x(t) \hat{\sigma}^x_0 \hat{\sigma}^x_l(t) \hat{\sigma}^x_0 \rangle  = 1$ and $C_{xx}(l,t) = 0$. In case, when $l = 1$
\begin{align*}
\hat{\sigma}^x_1(t) \hat{\sigma}_0^x \hat{\sigma}^x_1(t) \hat{\sigma}^x_0 &= \sum_{ij} e^{ \pm i (n\epsilon_i- m\bar{\epsilon}_i) } e^{ \pm i (n\epsilon_j- m\bar{\epsilon}_j) }|i_0 \bar{i}_1  \ldots \rangle  \langle i_0 i_1 \ldots | \hat{\sigma}^x_0   |j_0 \bar{j}_1 \ldots \rangle   \langle j_0 j_1 \ldots | \hat{\sigma}^x_0 \\
&= \sum_{ij} e^{ \pm i (n\epsilon_i- m\bar{\epsilon}_i) } e^{ \pm i (n\epsilon_j- m\bar{\epsilon}_j) }|i_0 \bar{i}_1  \ldots \rangle   \langle i_0 i_1 \ldots   | \bar{j}_0 \bar{j}_1 \ldots \rangle   \langle \bar{j}_0 j_1  \ldots |  \\
&= \sum_j e^{\pm iJ_0t\Delta j_0 \Delta j_1 }  |  j_0 j_1  \ldots  \rangle  \langle j_0 j_1  \ldots |
\end{align*}
\begin{figure*}
	\centering
	\includegraphics[width = 0.925\linewidth]{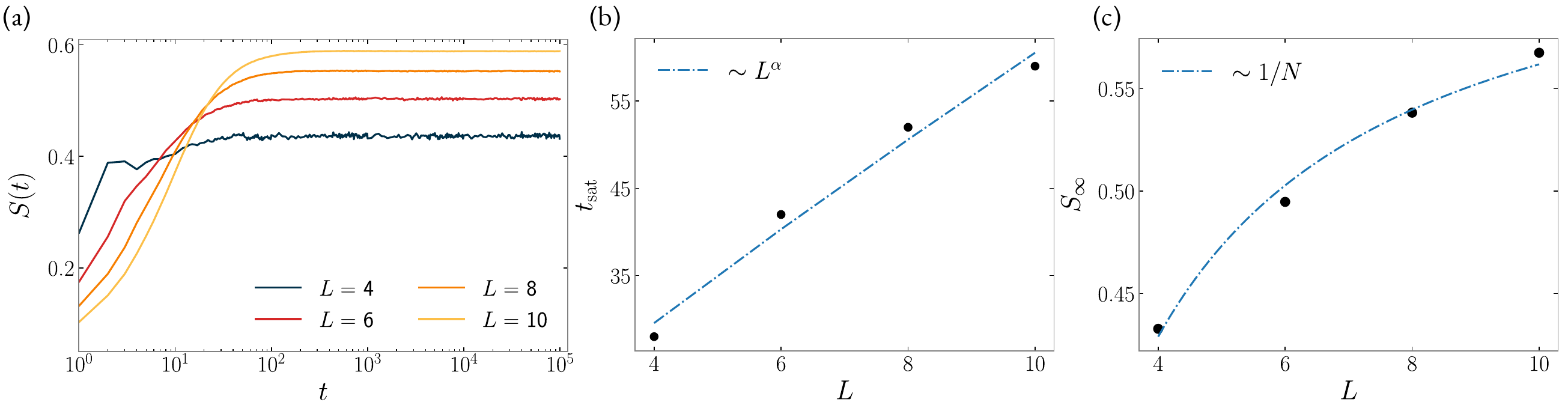}
	\caption{(a) The entanglement entropy $S(t)$ in melted FTC phase for the varying system-sizes $L$, and parameter values $J=1 =h^z$, $h^x = 0.5$ and the kicking phase $\phi = \pi/2 + 0.4$. (b) The saturation time plotted as function of system-size $L$, where the dash-dotted lines shows the fitted curve $\sim L^\alpha$ ($\alpha \approx 0.87$). (c) The saturation value $S(\infty)$ as a function of $L$ shows that the finite-size effect decays with system-size. The dash-dotted lines show the fitted curve $S^{(N\rightarrow \infty)}(\infty) + a/L$ with $S^{(N\rightarrow \infty)}(\infty) \approx 0.65$.}
	\label{fig:entropy melting}
\end{figure*}
which is diagonal operator with diagonal elements as phases. Therefore, we can write
\begin{align*}
C_{xx}(1,t) &= 1 - \mathfrak{Re } F_{xx}(1,t) = 1 - \langle \mathfrak{Re} ( \hat{\sigma}^x_1(t) \hat{\sigma}_0^x \hat{\sigma}^x_1(t) \hat{\sigma}^x_0) \rangle \\
&= 1 - \sum_j \cos (J_0 t \Delta j_0 \Delta j_1) \langle \Psi  |  j_0 j_1  \ldots  \rangle  \langle j_0 j_1  \ldots |  \Psi\rangle \\
&= 1 - \cos (4J_0 t ) \numberthis 
\end{align*}

since the coupling are chosen randomly, under disorder average 
\begin{equation}
\langle C_{xx}(1,t)\rangle = 1- \frac{\cos(6Jt)\sin(2Jt)}{2t}
\end{equation}
implying that $C_{xx}(1,t) \approx 1$ as $t\rightarrow \infty$. The $zx$ OTOC follows from evolution of $\hat{\sigma}^x_l$ and $\hat{\sigma}^z_l$ already computed $(l>0)$
\begin{align*}
\hat{\sigma}^z_l(t) \hat{\sigma}^x_0 \hat{\sigma}^z_l(t) \hat{\sigma}^x_0 &= (-1)^t \hat{\sigma}^z_l \hat{\sigma}^x_0 (-1)^t \hat{\sigma}_l^z \hat{\sigma}^x_0 = \hat{\sigma}^z_l \hat{\sigma}^x_0 \hat{\sigma}^z_l \hat{\sigma}^x_0 = I
\end{align*}
therefore, $C_{zx}(l,t) = 0$ $\forall \ t$.
\end{widetext}

\section{Ergodic phase}\label{sec:analysis in decoherence phase}
In this section, we examine the dynamics in the melted FTC or ergodic phase by introducing a phase kick $\phi$ far away from $\pi$. For our numerical simulations, we set the coupling parameters to $J = h^z = 1$, $h^x = 0.5$, and the $\phi = \pi/2 + 0.4$ for a system size of $L = 8$. Under these parameters, the mean level spacing ratio $r$ of the quasi-Hamiltonian $i \ln \hat{U}_F$ (associated with the unitary evolution in Eq.~\eqref{eq.UF}) is computed to be $\approx 0.53$ indicating a Wigner-Dyson distribution, typical of ergodic systems. As shown in Fig.~\ref{fig:FTC Melting} (left), the magnetization rapidly decays to zero due  to thermalization. Unlike the FTC phase, the melted phase lacks a preferred protected direction, leading to identical scrambling behaviour along all directions. This is evident in Fig.~\ref{fig:FTC Melting} (center and right), where the OTOCs $C_{zz}(l,t)$ and $C_{xx}(l,t)$ exhibit identical behavior. Similar to thermal systems, we observe a rapid early growth followed by saturation. In early growth regime ($C_{\alpha\beta}(l,t)<<1$), the behavior obeys the previously proposed conjecture\,\cite{xu_accessing_2020}
\begin{equation}C_{\alpha\beta}(l,t) \sim \exp(-\lambda(l-v_Bt)^{1+p}/t^p)\,.\end{equation}

Here, $v_B$ is the butterfly velocity, and $p$ is the wave front broadening coefficient. In Fig.\,\ref{fig:entropy melting}a, we present the behavior of entanglement entropy as a function of system size $L$, keeping the coupling parameters fixed. The entanglement entropy initially grows before reaching saturation. Unlike the FTC phase, where the saturation time exhibits exponential scaling, in the melted FTC, it scales polynomialy with system size $t_\text{sat} \sim L^\alpha$(see Fig.\,\ref{fig:entropy melting}b). The saturation time is determined by first computing the saturation value \( S(\infty) \) and then identifying the earliest time \( t_\text{sat} \) at which $|S(t_\text{sat}) - S(\infty)| \leq \epsilon$
where \( \epsilon \) is a tolerance parameter set to \( 10^{-3} \) in our numerics. Due to finite-size effects, the saturation value $S(\infty)$ is not  constant for the considered system sizes, as expected in a thermal system. We perform a finite-size scaling analysis of these saturation values to show that the correction decays to zero as $L\rightarrow \infty$ (see Fig.\,\ref{fig:entropy melting}c).

\bibliography{biblio} 

\begin{thebibliography}{105}%
\makeatletter
\providecommand \@ifxundefined [1]{%
 \@ifx{#1\undefined}
}%
\providecommand \@ifnum [1]{%
 \ifnum #1\expandafter \@firstoftwo
 \else \expandafter \@secondoftwo
 \fi
}%
\providecommand \@ifx [1]{%
 \ifx #1\expandafter \@firstoftwo
 \else \expandafter \@secondoftwo
 \fi
}%
\providecommand \natexlab [1]{#1}%
\providecommand \enquote  [1]{``#1''}%
\providecommand \bibnamefont  [1]{#1}%
\providecommand \bibfnamefont [1]{#1}%
\providecommand \citenamefont [1]{#1}%
\providecommand \href@noop [0]{\@secondoftwo}%
\providecommand \href [0]{\begingroup \@sanitize@url \@href}%
\providecommand \@href[1]{\@@startlink{#1}\@@href}%
\providecommand \@@href[1]{\endgroup#1\@@endlink}%
\providecommand \@sanitize@url [0]{\catcode `\\12\catcode `\$12\catcode
  `\&12\catcode `\#12\catcode `\^12\catcode `\_12\catcode `\%12\relax}%
\providecommand \@@startlink[1]{}%
\providecommand \@@endlink[0]{}%
\providecommand \url  [0]{\begingroup\@sanitize@url \@url }%
\providecommand \@url [1]{\endgroup\@href {#1}{\urlprefix }}%
\providecommand \urlprefix  [0]{URL }%
\providecommand \Eprint [0]{\href }%
\providecommand \doibase [0]{https://doi.org/}%
\providecommand \selectlanguage [0]{\@gobble}%
\providecommand \bibinfo  [0]{\@secondoftwo}%
\providecommand \bibfield  [0]{\@secondoftwo}%
\providecommand \translation [1]{[#1]}%
\providecommand \BibitemOpen [0]{}%
\providecommand \bibitemStop [0]{}%
\providecommand \bibitemNoStop [0]{.\EOS\space}%
\providecommand \EOS [0]{\spacefactor3000\relax}%
\providecommand \BibitemShut  [1]{\csname bibitem#1\endcsname}%
\let\auto@bib@innerbib\@empty
\bibitem [{\citenamefont {Zaletel}\ \emph {et~al.}(2023)\citenamefont
  {Zaletel}, \citenamefont {Lukin}, \citenamefont {Monroe}, \citenamefont
  {Nayak}, \citenamefont {Wilczek},\ and\ \citenamefont
  {Yao}}]{zaletel_colloquium_2023}%
  \BibitemOpen
  \bibfield  {author} {\bibinfo {author} {\bibfnamefont {M.~P.}\ \bibnamefont
  {Zaletel}}, \bibinfo {author} {\bibfnamefont {M.}~\bibnamefont {Lukin}},
  \bibinfo {author} {\bibfnamefont {C.}~\bibnamefont {Monroe}}, \bibinfo
  {author} {\bibfnamefont {C.}~\bibnamefont {Nayak}}, \bibinfo {author}
  {\bibfnamefont {F.}~\bibnamefont {Wilczek}},\ and\ \bibinfo {author}
  {\bibfnamefont {N.~Y.}\ \bibnamefont {Yao}},\ }\bibfield  {title} {\bibinfo
  {title} {Colloquium: {Quantum} and classical discrete time crystals},\ }\href
  {https://doi.org/10.1103/RevModPhys.95.031001} {\bibfield  {journal}
  {\bibinfo  {journal} {Rev. Mod. Phys.}\ }\textbf {\bibinfo {volume} {95}},\
  \bibinfo {pages} {031001} (\bibinfo {year} {2023})}\BibitemShut {NoStop}%
\bibitem [{\citenamefont {Else}\ \emph {et~al.}(2016)\citenamefont {Else},
  \citenamefont {Bauer},\ and\ \citenamefont {Nayak}}]{PhysRevLett.117.090402}%
  \BibitemOpen
  \bibfield  {author} {\bibinfo {author} {\bibfnamefont {D.~V.}\ \bibnamefont
  {Else}}, \bibinfo {author} {\bibfnamefont {B.}~\bibnamefont {Bauer}},\ and\
  \bibinfo {author} {\bibfnamefont {C.}~\bibnamefont {Nayak}},\ }\bibfield
  {title} {\bibinfo {title} {Floquet time crystals},\ }\href
  {https://doi.org/10.1103/PhysRevLett.117.090402} {\bibfield  {journal}
  {\bibinfo  {journal} {Phys. Rev. Lett.}\ }\textbf {\bibinfo {volume} {117}},\
  \bibinfo {pages} {090402} (\bibinfo {year} {2016})}\BibitemShut {NoStop}%
\bibitem [{\citenamefont {von Keyserlingk}\ and\ \citenamefont
  {Sondhi}(2016)}]{von_keyserlingk_phase_2016}%
  \BibitemOpen
  \bibfield  {author} {\bibinfo {author} {\bibfnamefont {C.~W.}\ \bibnamefont
  {von Keyserlingk}}\ and\ \bibinfo {author} {\bibfnamefont {S.~L.}\
  \bibnamefont {Sondhi}},\ }\bibfield  {title} {\bibinfo {title} {Phase
  structure of one-dimensional interacting {Floquet} systems. {II}.
  {Symmetry}-broken phases},\ }\href
  {https://doi.org/10.1103/PhysRevB.93.245146} {\bibfield  {journal} {\bibinfo
  {journal} {Phys. Rev. B}\ }\textbf {\bibinfo {volume} {93}},\ \bibinfo
  {pages} {245146} (\bibinfo {year} {2016})}\BibitemShut {NoStop}%
\bibitem [{\citenamefont {Khemani}\ \emph {et~al.}(2016)\citenamefont
  {Khemani}, \citenamefont {Lazarides}, \citenamefont {Moessner},\ and\
  \citenamefont {Sondhi}}]{khemani_phase_2016}%
  \BibitemOpen
  \bibfield  {author} {\bibinfo {author} {\bibfnamefont {V.}~\bibnamefont
  {Khemani}}, \bibinfo {author} {\bibfnamefont {A.}~\bibnamefont {Lazarides}},
  \bibinfo {author} {\bibfnamefont {R.}~\bibnamefont {Moessner}},\ and\
  \bibinfo {author} {\bibfnamefont {S.}~\bibnamefont {Sondhi}},\ }\bibfield
  {title} {{\selectlanguage {english}\bibinfo {title} {Phase {Structure} of
  {Driven} {Quantum} {Systems}}},\ }\href
  {https://doi.org/10.1103/PhysRevLett.116.250401} {\bibfield  {journal}
  {\bibinfo  {journal} {Phys. Rev. Lett.}\ }\textbf {\bibinfo {volume} {116}},\
  \bibinfo {pages} {250401} (\bibinfo {year} {2016})}\BibitemShut {NoStop}%
\bibitem [{\citenamefont {Yao}\ \emph {et~al.}(2017)\citenamefont {Yao},
  \citenamefont {Potter}, \citenamefont {Potirniche},\ and\ \citenamefont
  {Vishwanath}}]{yao_discrete_2017}%
  \BibitemOpen
  \bibfield  {author} {\bibinfo {author} {\bibfnamefont {N.}~\bibnamefont
  {Yao}}, \bibinfo {author} {\bibfnamefont {A.}~\bibnamefont {Potter}},
  \bibinfo {author} {\bibfnamefont {I.-D.}\ \bibnamefont {Potirniche}},\ and\
  \bibinfo {author} {\bibfnamefont {A.}~\bibnamefont {Vishwanath}},\ }\bibfield
   {title} {{\selectlanguage {english}\bibinfo {title} {Discrete {Time}
  {Crystals}: {Rigidity}, {Criticality}, and {Realizations}}},\ }\href
  {https://doi.org/10.1103/PhysRevLett.118.030401} {\bibfield  {journal}
  {\bibinfo  {journal} {Phys. Rev. Lett.}\ }\textbf {\bibinfo {volume} {118}},\
  \bibinfo {pages} {030401} (\bibinfo {year} {2017})}\BibitemShut {NoStop}%
\bibitem [{\citenamefont {Russomanno}\ \emph {et~al.}(2017)\citenamefont
  {Russomanno}, \citenamefont {Iemini}, \citenamefont {Dalmonte},\ and\
  \citenamefont {Fazio}}]{russomanno_floquet_2017}%
  \BibitemOpen
  \bibfield  {author} {\bibinfo {author} {\bibfnamefont {A.}~\bibnamefont
  {Russomanno}}, \bibinfo {author} {\bibfnamefont {F.}~\bibnamefont {Iemini}},
  \bibinfo {author} {\bibfnamefont {M.}~\bibnamefont {Dalmonte}},\ and\
  \bibinfo {author} {\bibfnamefont {R.}~\bibnamefont {Fazio}},\ }\bibfield
  {title} {\bibinfo {title} {Floquet time crystal in the
  {Lipkin}-{Meshkov}-{Glick} model},\ }\href
  {https://doi.org/10.1103/PhysRevB.95.214307} {\bibfield  {journal} {\bibinfo
  {journal} {Phys. Rev. B}\ }\textbf {\bibinfo {volume} {95}},\ \bibinfo
  {pages} {214307} (\bibinfo {year} {2017})}\BibitemShut {NoStop}%
\bibitem [{\citenamefont {Surace}\ \emph {et~al.}(2019)\citenamefont {Surace},
  \citenamefont {Russomanno}, \citenamefont {Dalmonte}, \citenamefont {Silva},
  \citenamefont {Fazio},\ and\ \citenamefont {Iemini}}]{surace_floquet_2019}%
  \BibitemOpen
  \bibfield  {author} {\bibinfo {author} {\bibfnamefont {F.~M.}\ \bibnamefont
  {Surace}}, \bibinfo {author} {\bibfnamefont {A.}~\bibnamefont {Russomanno}},
  \bibinfo {author} {\bibfnamefont {M.}~\bibnamefont {Dalmonte}}, \bibinfo
  {author} {\bibfnamefont {A.}~\bibnamefont {Silva}}, \bibinfo {author}
  {\bibfnamefont {R.}~\bibnamefont {Fazio}},\ and\ \bibinfo {author}
  {\bibfnamefont {F.}~\bibnamefont {Iemini}},\ }\bibfield  {title} {\bibinfo
  {title} {Floquet time crystals in clock models},\ }\href
  {https://doi.org/10.1103/PhysRevB.99.104303} {\bibfield  {journal} {\bibinfo
  {journal} {Phys. Rev. B}\ }\textbf {\bibinfo {volume} {99}},\ \bibinfo
  {pages} {104303} (\bibinfo {year} {2019})}\BibitemShut {NoStop}%
\bibitem [{\citenamefont {Yang}\ and\ \citenamefont
  {Cai}(2021)}]{yang_dynamical_2021}%
  \BibitemOpen
  \bibfield  {author} {\bibinfo {author} {\bibfnamefont {X.}~\bibnamefont
  {Yang}}\ and\ \bibinfo {author} {\bibfnamefont {Z.}~\bibnamefont {Cai}},\
  }\bibfield  {title} {\bibinfo {title} {Dynamical {Transitions} and {Critical}
  {Behavior} between {Discrete} {Time} {Crystal} {Phases}},\ }\href
  {https://doi.org/10.1103/PhysRevLett.126.020602} {\bibfield  {journal}
  {\bibinfo  {journal} {Phys. Rev. Lett.}\ }\textbf {\bibinfo {volume} {126}},\
  \bibinfo {pages} {020602} (\bibinfo {year} {2021})}\BibitemShut {NoStop}%
\bibitem [{\citenamefont {Morita}\ and\ \citenamefont
  {Kaneko}(2006)}]{morita_collective_2006}%
  \BibitemOpen
  \bibfield  {author} {\bibinfo {author} {\bibfnamefont {H.}~\bibnamefont
  {Morita}}\ and\ \bibinfo {author} {\bibfnamefont {K.}~\bibnamefont
  {Kaneko}},\ }\bibfield  {title} {\bibinfo {title} {Collective {Oscillation}
  in a {Hamiltonian} {System}},\ }\href
  {https://doi.org/10.1103/PhysRevLett.96.050602} {\bibfield  {journal}
  {\bibinfo  {journal} {Phys. Rev. Lett.}\ }\textbf {\bibinfo {volume} {96}},\
  \bibinfo {pages} {050602} (\bibinfo {year} {2006})}\BibitemShut {NoStop}%
\bibitem [{\citenamefont {Ojeda~Collado}\ \emph {et~al.}(2021)\citenamefont
  {Ojeda~Collado}, \citenamefont {Usaj}, \citenamefont {Balseiro},
  \citenamefont {Zanette},\ and\ \citenamefont
  {Lorenzana}}]{ojeda_collado_emergent_2021}%
  \BibitemOpen
  \bibfield  {author} {\bibinfo {author} {\bibfnamefont {H.~P.}\ \bibnamefont
  {Ojeda~Collado}}, \bibinfo {author} {\bibfnamefont {G.}~\bibnamefont {Usaj}},
  \bibinfo {author} {\bibfnamefont {C.~A.}\ \bibnamefont {Balseiro}}, \bibinfo
  {author} {\bibfnamefont {D.~H.}\ \bibnamefont {Zanette}},\ and\ \bibinfo
  {author} {\bibfnamefont {J.}~\bibnamefont {Lorenzana}},\ }\bibfield  {title}
  {\bibinfo {title} {Emergent parametric resonances and time-crystal phases in
  driven {Bardeen}-{Cooper}-{Schrieffer} systems},\ }\href
  {https://doi.org/10.1103/PhysRevResearch.3.L042023} {\bibfield  {journal}
  {\bibinfo  {journal} {Phys. Rev. Res.}\ }\textbf {\bibinfo {volume} {3}},\
  \bibinfo {pages} {L042023} (\bibinfo {year} {2021})}\BibitemShut {NoStop}%
\bibitem [{\citenamefont {Nurwantoro}\ \emph {et~al.}(2019)\citenamefont
  {Nurwantoro}, \citenamefont {Bomantara},\ and\ \citenamefont
  {Gong}}]{nurwantoro_discrete_2019}%
  \BibitemOpen
  \bibfield  {author} {\bibinfo {author} {\bibfnamefont {P.}~\bibnamefont
  {Nurwantoro}}, \bibinfo {author} {\bibfnamefont {R.~W.}\ \bibnamefont
  {Bomantara}},\ and\ \bibinfo {author} {\bibfnamefont {J.}~\bibnamefont
  {Gong}},\ }\bibfield  {title} {\bibinfo {title} {Discrete time crystals in
  many-body quantum chaos},\ }\href
  {https://doi.org/10.1103/PhysRevB.100.214311} {\bibfield  {journal} {\bibinfo
   {journal} {Phys. Rev. B}\ }\textbf {\bibinfo {volume} {100}},\ \bibinfo
  {pages} {214311} (\bibinfo {year} {2019})}\BibitemShut {NoStop}%
\bibitem [{\citenamefont {Pizzi}\ \emph {et~al.}(2021)\citenamefont {Pizzi},
  \citenamefont {Knolle},\ and\ \citenamefont
  {Nunnenkamp}}]{pizzi_higher-order_2021}%
  \BibitemOpen
  \bibfield  {author} {\bibinfo {author} {\bibfnamefont {A.}~\bibnamefont
  {Pizzi}}, \bibinfo {author} {\bibfnamefont {J.}~\bibnamefont {Knolle}},\ and\
  \bibinfo {author} {\bibfnamefont {A.}~\bibnamefont {Nunnenkamp}},\ }\bibfield
   {title} {{\selectlanguage {english}\bibinfo {title} {Higher-order and
  fractional discrete time crystals in clean long-range interacting systems}},\
  }\href {https://doi.org/10.1038/s41467-021-22583-5} {\bibfield  {journal}
  {\bibinfo  {journal} {Nature Communications}\ }\textbf {\bibinfo {volume}
  {12}},\ \bibinfo {pages} {2341} (\bibinfo {year} {2021})}\BibitemShut
  {NoStop}%
\bibitem [{\citenamefont {Muñoz-Arias}\ \emph {et~al.}(2022)\citenamefont
  {Muñoz-Arias}, \citenamefont {Chinni},\ and\ \citenamefont
  {Poggi}}]{munoz_arias_floquet_2022}%
  \BibitemOpen
  \bibfield  {author} {\bibinfo {author} {\bibfnamefont {M.~H.}\ \bibnamefont
  {Muñoz-Arias}}, \bibinfo {author} {\bibfnamefont {K.}~\bibnamefont
  {Chinni}},\ and\ \bibinfo {author} {\bibfnamefont {P.~M.}\ \bibnamefont
  {Poggi}},\ }\bibfield  {title} {\bibinfo {title} {Floquet time crystals in
  driven spin systems with all-to-all \$p\$-body interactions},\ }\href
  {https://doi.org/10.1103/PhysRevResearch.4.023018} {\bibfield  {journal}
  {\bibinfo  {journal} {Phys. Rev. Res.}\ }\textbf {\bibinfo {volume} {4}},\
  \bibinfo {pages} {023018} (\bibinfo {year} {2022})}\BibitemShut {NoStop}%
\bibitem [{\citenamefont {Giachetti}\ \emph {et~al.}(2023)\citenamefont
  {Giachetti}, \citenamefont {Solfanelli}, \citenamefont {Correale},\ and\
  \citenamefont {Defenu}}]{giachetti_fractal_2023}%
  \BibitemOpen
  \bibfield  {author} {\bibinfo {author} {\bibfnamefont {G.}~\bibnamefont
  {Giachetti}}, \bibinfo {author} {\bibfnamefont {A.}~\bibnamefont
  {Solfanelli}}, \bibinfo {author} {\bibfnamefont {L.}~\bibnamefont
  {Correale}},\ and\ \bibinfo {author} {\bibfnamefont {N.}~\bibnamefont
  {Defenu}},\ }\bibfield  {title} {\bibinfo {title} {Fractal nature of
  high-order time crystal phases},\ }\href
  {https://doi.org/10.1103/PhysRevB.108.L140102} {\bibfield  {journal}
  {\bibinfo  {journal} {Phys. Rev. B}\ }\textbf {\bibinfo {volume} {108}},\
  \bibinfo {pages} {L140102} (\bibinfo {year} {2023})}\BibitemShut {NoStop}%
\bibitem [{\citenamefont {Liu}\ \emph {et~al.}(2023{\natexlab{a}})\citenamefont
  {Liu}, \citenamefont {Zhang}, \citenamefont {Hsieh}, \citenamefont {Zhang},\
  and\ \citenamefont {Yao}}]{liu_discrete_2023}%
  \BibitemOpen
  \bibfield  {author} {\bibinfo {author} {\bibfnamefont {S.}~\bibnamefont
  {Liu}}, \bibinfo {author} {\bibfnamefont {S.-X.}\ \bibnamefont {Zhang}},
  \bibinfo {author} {\bibfnamefont {C.-Y.}\ \bibnamefont {Hsieh}}, \bibinfo
  {author} {\bibfnamefont {S.}~\bibnamefont {Zhang}},\ and\ \bibinfo {author}
  {\bibfnamefont {H.}~\bibnamefont {Yao}},\ }\bibfield  {title} {\bibinfo
  {title} {Discrete {Time} {Crystal} {Enabled} by {Stark} {Many}-{Body}
  {Localization}},\ }\href {https://doi.org/10.1103/PhysRevLett.130.120403}
  {\bibfield  {journal} {\bibinfo  {journal} {Phys. Rev. Lett.}\ }\textbf
  {\bibinfo {volume} {130}},\ \bibinfo {pages} {120403} (\bibinfo {year}
  {2023}{\natexlab{a}})}\BibitemShut {NoStop}%
\bibitem [{\citenamefont {Lev}\ and\ \citenamefont
  {Lazarides}(2024)}]{lev_discrete_2024}%
  \BibitemOpen
  \bibfield  {author} {\bibinfo {author} {\bibfnamefont {Y.~B.}\ \bibnamefont
  {Lev}}\ and\ \bibinfo {author} {\bibfnamefont {A.}~\bibnamefont
  {Lazarides}},\ }\href@noop {} {\bibinfo {title} {{Discrete time-crystals in
  linear potentials}}} (\bibinfo {year} {2024}),\ \Eprint
  {https://arxiv.org/abs/2403.01912} {arXiv:2403.01912 [cond-mat.dis-nn]}
  \BibitemShut {NoStop}%
\bibitem [{\citenamefont {Maskara}\ \emph
  {et~al.}(2021{\natexlab{a}})\citenamefont {Maskara}, \citenamefont
  {Michailidis}, \citenamefont {Ho}, \citenamefont {Bluvstein}, \citenamefont
  {Choi}, \citenamefont {Lukin},\ and\ \citenamefont
  {Serbyn}}]{maskara_discrete_2021}%
  \BibitemOpen
  \bibfield  {author} {\bibinfo {author} {\bibfnamefont {N.}~\bibnamefont
  {Maskara}}, \bibinfo {author} {\bibfnamefont {A.}~\bibnamefont
  {Michailidis}}, \bibinfo {author} {\bibfnamefont {W.}~\bibnamefont {Ho}},
  \bibinfo {author} {\bibfnamefont {D.}~\bibnamefont {Bluvstein}}, \bibinfo
  {author} {\bibfnamefont {S.}~\bibnamefont {Choi}}, \bibinfo {author}
  {\bibfnamefont {M.}~\bibnamefont {Lukin}},\ and\ \bibinfo {author}
  {\bibfnamefont {M.}~\bibnamefont {Serbyn}},\ }\bibfield  {title} {\bibinfo
  {title} {Discrete {Time}-{Crystalline} {Order} {Enabled} by {Quantum}
  {Many}-{Body} {Scars}: {Entanglement} {Steering} via {Periodic} {Driving}},\
  }\href {https://doi.org/10.1103/PhysRevLett.127.090602} {\bibfield  {journal}
  {\bibinfo  {journal} {Phys. Rev. Lett.}\ }\textbf {\bibinfo {volume} {127}},\
  \bibinfo {pages} {090602} (\bibinfo {year} {2021}{\natexlab{a}})}\BibitemShut
  {NoStop}%
\bibitem [{\citenamefont {Huang}(2023)}]{huang_analytical_2023}%
  \BibitemOpen
  \bibfield  {author} {\bibinfo {author} {\bibfnamefont {B.}~\bibnamefont
  {Huang}},\ }\bibfield  {title} {\bibinfo {title} {Analytical theory of cat
  scars with discrete time-crystalline dynamics in {Floquet} systems},\ }\href
  {https://doi.org/10.1103/PhysRevB.108.104309} {\bibfield  {journal} {\bibinfo
   {journal} {Phys. Rev. B}\ }\textbf {\bibinfo {volume} {108}},\ \bibinfo
  {pages} {104309} (\bibinfo {year} {2023})}\BibitemShut {NoStop}%
\bibitem [{\citenamefont {Iemini}\ \emph {et~al.}(2018)\citenamefont {Iemini},
  \citenamefont {Russomanno}, \citenamefont {Keeling}, \citenamefont
  {Schir\`o}, \citenamefont {Dalmonte},\ and\ \citenamefont
  {Fazio}}]{PhysRevLett.121.035301}%
  \BibitemOpen
  \bibfield  {author} {\bibinfo {author} {\bibfnamefont {F.}~\bibnamefont
  {Iemini}}, \bibinfo {author} {\bibfnamefont {A.}~\bibnamefont {Russomanno}},
  \bibinfo {author} {\bibfnamefont {J.}~\bibnamefont {Keeling}}, \bibinfo
  {author} {\bibfnamefont {M.}~\bibnamefont {Schir\`o}}, \bibinfo {author}
  {\bibfnamefont {M.}~\bibnamefont {Dalmonte}},\ and\ \bibinfo {author}
  {\bibfnamefont {R.}~\bibnamefont {Fazio}},\ }\bibfield  {title} {\bibinfo
  {title} {Boundary time crystals},\ }\href
  {https://doi.org/10.1103/PhysRevLett.121.035301} {\bibfield  {journal}
  {\bibinfo  {journal} {Phys. Rev. Lett.}\ }\textbf {\bibinfo {volume} {121}},\
  \bibinfo {pages} {035301} (\bibinfo {year} {2018})}\BibitemShut {NoStop}%
\bibitem [{\citenamefont {Souza}\ \emph {et~al.}(2023)\citenamefont {Souza},
  \citenamefont {dos Prazeres},\ and\ \citenamefont
  {Iemini}}]{PhysRevLett.130.180401}%
  \BibitemOpen
  \bibfield  {author} {\bibinfo {author} {\bibfnamefont {L.~d.~S.}\
  \bibnamefont {Souza}}, \bibinfo {author} {\bibfnamefont {L.~F.}\ \bibnamefont
  {dos Prazeres}},\ and\ \bibinfo {author} {\bibfnamefont {F.}~\bibnamefont
  {Iemini}},\ }\bibfield  {title} {\bibinfo {title} {Sufficient condition for
  gapless spin-boson lindbladians, and its connection to dissipative time
  crystals},\ }\href {https://doi.org/10.1103/PhysRevLett.130.180401}
  {\bibfield  {journal} {\bibinfo  {journal} {Phys. Rev. Lett.}\ }\textbf
  {\bibinfo {volume} {130}},\ \bibinfo {pages} {180401} (\bibinfo {year}
  {2023})}\BibitemShut {NoStop}%
\bibitem [{\citenamefont {Kongkhambut}\ \emph {et~al.}(2022)\citenamefont
  {Kongkhambut}, \citenamefont {Skulte}, \citenamefont {Mathey}, \citenamefont
  {Cosme}, \citenamefont {Hemmerich},\ and\ \citenamefont
  {Keßler}}]{Phatthamon2022}%
  \BibitemOpen
  \bibfield  {author} {\bibinfo {author} {\bibfnamefont {P.}~\bibnamefont
  {Kongkhambut}}, \bibinfo {author} {\bibfnamefont {J.}~\bibnamefont {Skulte}},
  \bibinfo {author} {\bibfnamefont {L.}~\bibnamefont {Mathey}}, \bibinfo
  {author} {\bibfnamefont {J.~G.}\ \bibnamefont {Cosme}}, \bibinfo {author}
  {\bibfnamefont {A.}~\bibnamefont {Hemmerich}},\ and\ \bibinfo {author}
  {\bibfnamefont {H.}~\bibnamefont {Keßler}},\ }\bibfield  {title} {\bibinfo
  {title} {Observation of a continuous time crystal},\ }\href
  {https://doi.org/10.1126/science.abo3382} {\bibfield  {journal} {\bibinfo
  {journal} {Science}\ }\textbf {\bibinfo {volume} {377}},\ \bibinfo {pages}
  {670} (\bibinfo {year} {2022})}\BibitemShut {NoStop}%
\bibitem [{\citenamefont {Nakanishi}\ and\ \citenamefont
  {Sasamoto}(2023)}]{PhysRevA.107.L010201}%
  \BibitemOpen
  \bibfield  {author} {\bibinfo {author} {\bibfnamefont {Y.}~\bibnamefont
  {Nakanishi}}\ and\ \bibinfo {author} {\bibfnamefont {T.}~\bibnamefont
  {Sasamoto}},\ }\bibfield  {title} {\bibinfo {title} {Dissipative time
  crystals originating from parity-time symmetry},\ }\href
  {https://doi.org/10.1103/PhysRevA.107.L010201} {\bibfield  {journal}
  {\bibinfo  {journal} {Phys. Rev. A}\ }\textbf {\bibinfo {volume} {107}},\
  \bibinfo {pages} {L010201} (\bibinfo {year} {2023})}\BibitemShut {NoStop}%
\bibitem [{\citenamefont {Krishna}\ \emph {et~al.}(2023)\citenamefont
  {Krishna}, \citenamefont {Solanki}, \citenamefont
  {Hajdu\ifmmode~\check{s}\else \v{s}\fi{}ek},\ and\ \citenamefont
  {Vinjanampathy}}]{PhysRevLett.130.150401}%
  \BibitemOpen
  \bibfield  {author} {\bibinfo {author} {\bibfnamefont {M.}~\bibnamefont
  {Krishna}}, \bibinfo {author} {\bibfnamefont {P.}~\bibnamefont {Solanki}},
  \bibinfo {author} {\bibfnamefont {M.}~\bibnamefont
  {Hajdu\ifmmode~\check{s}\else \v{s}\fi{}ek}},\ and\ \bibinfo {author}
  {\bibfnamefont {S.}~\bibnamefont {Vinjanampathy}},\ }\bibfield  {title}
  {\bibinfo {title} {Measurement-induced continuous time crystals},\ }\href
  {https://doi.org/10.1103/PhysRevLett.130.150401} {\bibfield  {journal}
  {\bibinfo  {journal} {Phys. Rev. Lett.}\ }\textbf {\bibinfo {volume} {130}},\
  \bibinfo {pages} {150401} (\bibinfo {year} {2023})}\BibitemShut {NoStop}%
\bibitem [{\citenamefont {Li}\ \emph {et~al.}(2023)\citenamefont {Li},
  \citenamefont {Li},\ and\ \citenamefont {Jin}}]{PhysRevA.107.032219}%
  \BibitemOpen
  \bibfield  {author} {\bibinfo {author} {\bibfnamefont {X.}~\bibnamefont
  {Li}}, \bibinfo {author} {\bibfnamefont {Y.}~\bibnamefont {Li}},\ and\
  \bibinfo {author} {\bibfnamefont {J.}~\bibnamefont {Jin}},\ }\bibfield
  {title} {\bibinfo {title} {Synchronization of persistent oscillations in spin
  systems with nonlocal dissipation},\ }\href
  {https://doi.org/10.1103/PhysRevA.107.032219} {\bibfield  {journal} {\bibinfo
   {journal} {Phys. Rev. A}\ }\textbf {\bibinfo {volume} {107}},\ \bibinfo
  {pages} {032219} (\bibinfo {year} {2023})}\BibitemShut {NoStop}%
\bibitem [{\citenamefont {Bu\ifmmode~\check{c}\else \v{c}\fi{}a}\ and\
  \citenamefont {Jaksch}(2019)}]{PhysRevLett.123.260401}%
  \BibitemOpen
  \bibfield  {author} {\bibinfo {author} {\bibfnamefont {B.}~\bibnamefont
  {Bu\ifmmode~\check{c}\else \v{c}\fi{}a}}\ and\ \bibinfo {author}
  {\bibfnamefont {D.}~\bibnamefont {Jaksch}},\ }\bibfield  {title} {\bibinfo
  {title} {Dissipation induced nonstationarity in a quantum gas},\ }\href
  {https://doi.org/10.1103/PhysRevLett.123.260401} {\bibfield  {journal}
  {\bibinfo  {journal} {Phys. Rev. Lett.}\ }\textbf {\bibinfo {volume} {123}},\
  \bibinfo {pages} {260401} (\bibinfo {year} {2019})}\BibitemShut {NoStop}%
\bibitem [{\citenamefont {Zhu}\ \emph {et~al.}(2019)\citenamefont {Zhu},
  \citenamefont {Marino}, \citenamefont {Yao}, \citenamefont {Lukin},\ and\
  \citenamefont {Demler}}]{Zhu_2019}%
  \BibitemOpen
  \bibfield  {author} {\bibinfo {author} {\bibfnamefont {B.}~\bibnamefont
  {Zhu}}, \bibinfo {author} {\bibfnamefont {J.}~\bibnamefont {Marino}},
  \bibinfo {author} {\bibfnamefont {N.~Y.}\ \bibnamefont {Yao}}, \bibinfo
  {author} {\bibfnamefont {M.~D.}\ \bibnamefont {Lukin}},\ and\ \bibinfo
  {author} {\bibfnamefont {E.~A.}\ \bibnamefont {Demler}},\ }\bibfield  {title}
  {\bibinfo {title} {Dicke time crystals in driven-dissipative quantum
  many-body systems},\ }\href {https://doi.org/10.1088/1367-2630/ab2afe}
  {\bibfield  {journal} {\bibinfo  {journal} {New Journal of Physics}\ }\textbf
  {\bibinfo {volume} {21}},\ \bibinfo {pages} {073028} (\bibinfo {year}
  {2019})}\BibitemShut {NoStop}%
\bibitem [{\citenamefont {Gambetta}\ \emph {et~al.}(2019)\citenamefont
  {Gambetta}, \citenamefont {Carollo}, \citenamefont {Marcuzzi}, \citenamefont
  {Garrahan},\ and\ \citenamefont {Lesanovsky}}]{PhysRevLett.122.015701}%
  \BibitemOpen
  \bibfield  {author} {\bibinfo {author} {\bibfnamefont {F.~M.}\ \bibnamefont
  {Gambetta}}, \bibinfo {author} {\bibfnamefont {F.}~\bibnamefont {Carollo}},
  \bibinfo {author} {\bibfnamefont {M.}~\bibnamefont {Marcuzzi}}, \bibinfo
  {author} {\bibfnamefont {J.~P.}\ \bibnamefont {Garrahan}},\ and\ \bibinfo
  {author} {\bibfnamefont {I.}~\bibnamefont {Lesanovsky}},\ }\bibfield  {title}
  {\bibinfo {title} {Discrete time crystals in the absence of manifest
  symmetries or disorder in open quantum systems},\ }\href
  {https://doi.org/10.1103/PhysRevLett.122.015701} {\bibfield  {journal}
  {\bibinfo  {journal} {Phys. Rev. Lett.}\ }\textbf {\bibinfo {volume} {122}},\
  \bibinfo {pages} {015701} (\bibinfo {year} {2019})}\BibitemShut {NoStop}%
\bibitem [{\citenamefont {Tucker}\ \emph {et~al.}(2018)\citenamefont {Tucker},
  \citenamefont {Zhu}, \citenamefont {Lewis-Swan}, \citenamefont {Marino},
  \citenamefont {Jimenez}, \citenamefont {Restrepo},\ and\ \citenamefont
  {Rey}}]{Tucker_2018}%
  \BibitemOpen
  \bibfield  {author} {\bibinfo {author} {\bibfnamefont {K.}~\bibnamefont
  {Tucker}}, \bibinfo {author} {\bibfnamefont {B.}~\bibnamefont {Zhu}},
  \bibinfo {author} {\bibfnamefont {R.~J.}\ \bibnamefont {Lewis-Swan}},
  \bibinfo {author} {\bibfnamefont {J.}~\bibnamefont {Marino}}, \bibinfo
  {author} {\bibfnamefont {F.}~\bibnamefont {Jimenez}}, \bibinfo {author}
  {\bibfnamefont {J.~G.}\ \bibnamefont {Restrepo}},\ and\ \bibinfo {author}
  {\bibfnamefont {A.~M.}\ \bibnamefont {Rey}},\ }\bibfield  {title} {\bibinfo
  {title} {Shattered time: can a dissipative time crystal survive many-body
  correlations?},\ }\href {https://doi.org/10.1088/1367-2630/aaf18b} {\bibfield
   {journal} {\bibinfo  {journal} {New Journal of Physics}\ }\textbf {\bibinfo
  {volume} {20}},\ \bibinfo {pages} {123003} (\bibinfo {year}
  {2018})}\BibitemShut {NoStop}%
\bibitem [{\citenamefont {Wang}\ \emph {et~al.}(2018)\citenamefont {Wang},
  \citenamefont {Xing}, \citenamefont {Carlo},\ and\ \citenamefont
  {Poletti}}]{PhysRevE.97.020202}%
  \BibitemOpen
  \bibfield  {author} {\bibinfo {author} {\bibfnamefont {R.~R.~W.}\
  \bibnamefont {Wang}}, \bibinfo {author} {\bibfnamefont {B.}~\bibnamefont
  {Xing}}, \bibinfo {author} {\bibfnamefont {G.~G.}\ \bibnamefont {Carlo}},\
  and\ \bibinfo {author} {\bibfnamefont {D.}~\bibnamefont {Poletti}},\
  }\bibfield  {title} {\bibinfo {title} {Period doubling in period-one steady
  states},\ }\href {https://doi.org/10.1103/PhysRevE.97.020202} {\bibfield
  {journal} {\bibinfo  {journal} {Phys. Rev. E}\ }\textbf {\bibinfo {volume}
  {97}},\ \bibinfo {pages} {020202} (\bibinfo {year} {2018})}\BibitemShut
  {NoStop}%
\bibitem [{\citenamefont {Gong}\ \emph {et~al.}(2018)\citenamefont {Gong},
  \citenamefont {Hamazaki},\ and\ \citenamefont
  {Ueda}}]{PhysRevLett.120.040404}%
  \BibitemOpen
  \bibfield  {author} {\bibinfo {author} {\bibfnamefont {Z.}~\bibnamefont
  {Gong}}, \bibinfo {author} {\bibfnamefont {R.}~\bibnamefont {Hamazaki}},\
  and\ \bibinfo {author} {\bibfnamefont {M.}~\bibnamefont {Ueda}},\ }\bibfield
  {title} {\bibinfo {title} {Discrete time-crystalline order in cavity and
  circuit qed systems},\ }\href
  {https://doi.org/10.1103/PhysRevLett.120.040404} {\bibfield  {journal}
  {\bibinfo  {journal} {Phys. Rev. Lett.}\ }\textbf {\bibinfo {volume} {120}},\
  \bibinfo {pages} {040404} (\bibinfo {year} {2018})}\BibitemShut {NoStop}%
\bibitem [{\citenamefont {Kyprianidis}\ \emph {et~al.}(2021)\citenamefont
  {Kyprianidis}, \citenamefont {Machado}, \citenamefont {Morong}, \citenamefont
  {Becker}, \citenamefont {Collins}, \citenamefont {Else}, \citenamefont
  {Feng}, \citenamefont {Hess}, \citenamefont {Nayak}, \citenamefont {Pagano}
  \emph {et~al.}}]{kyprianidis_observation_2021}%
  \BibitemOpen
  \bibfield  {author} {\bibinfo {author} {\bibfnamefont {A.}~\bibnamefont
  {Kyprianidis}}, \bibinfo {author} {\bibfnamefont {F.}~\bibnamefont
  {Machado}}, \bibinfo {author} {\bibfnamefont {W.}~\bibnamefont {Morong}},
  \bibinfo {author} {\bibfnamefont {P.}~\bibnamefont {Becker}}, \bibinfo
  {author} {\bibfnamefont {K.~S.}\ \bibnamefont {Collins}}, \bibinfo {author}
  {\bibfnamefont {D.~V.}\ \bibnamefont {Else}}, \bibinfo {author}
  {\bibfnamefont {L.}~\bibnamefont {Feng}}, \bibinfo {author} {\bibfnamefont
  {P.~W.}\ \bibnamefont {Hess}}, \bibinfo {author} {\bibfnamefont
  {C.}~\bibnamefont {Nayak}}, \bibinfo {author} {\bibfnamefont
  {G.}~\bibnamefont {Pagano}}, \emph {et~al.},\ }\bibfield  {title} {\bibinfo
  {title} {Observation of a prethermal discrete time crystal},\ }\href
  {https://doi.org/10.1126/science.abg8102} {\bibfield  {journal} {\bibinfo
  {journal} {Science}\ }\textbf {\bibinfo {volume} {372}},\ \bibinfo {pages}
  {1192} (\bibinfo {year} {2021})}\BibitemShut {NoStop}%
\bibitem [{\citenamefont {Stasiuk}\ and\ \citenamefont
  {Cappellaro}(2023)}]{stasiuk_observation_2023}%
  \BibitemOpen
  \bibfield  {author} {\bibinfo {author} {\bibfnamefont {A.}~\bibnamefont
  {Stasiuk}}\ and\ \bibinfo {author} {\bibfnamefont {P.}~\bibnamefont
  {Cappellaro}},\ }\bibfield  {title} {\bibinfo {title} {Observation of a
  {Prethermal} \${U}(1)\$ {Discrete} {Time} {Crystal}},\ }\href
  {https://doi.org/10.1103/PhysRevX.13.041016} {\bibfield  {journal} {\bibinfo
  {journal} {Phys. Rev. X}\ }\textbf {\bibinfo {volume} {13}},\ \bibinfo
  {pages} {041016} (\bibinfo {year} {2023})}\BibitemShut {NoStop}%
\bibitem [{\citenamefont {Beatrez}\ \emph {et~al.}(2023)\citenamefont
  {Beatrez}, \citenamefont {Fleckenstein}, \citenamefont {Pillai},
  \citenamefont {de~Leon~Sanchez}, \citenamefont {Akkiraju}, \citenamefont
  {Diaz~Alcala}, \citenamefont {Conti}, \citenamefont {Reshetikhin},
  \citenamefont {Druga}, \citenamefont {Bukov} \emph
  {et~al.}}]{beatrez_critical_2023}%
  \BibitemOpen
  \bibfield  {author} {\bibinfo {author} {\bibfnamefont {W.}~\bibnamefont
  {Beatrez}}, \bibinfo {author} {\bibfnamefont {C.}~\bibnamefont
  {Fleckenstein}}, \bibinfo {author} {\bibfnamefont {A.}~\bibnamefont
  {Pillai}}, \bibinfo {author} {\bibfnamefont {E.}~\bibnamefont
  {de~Leon~Sanchez}}, \bibinfo {author} {\bibfnamefont {A.}~\bibnamefont
  {Akkiraju}}, \bibinfo {author} {\bibfnamefont {J.}~\bibnamefont
  {Diaz~Alcala}}, \bibinfo {author} {\bibfnamefont {S.}~\bibnamefont {Conti}},
  \bibinfo {author} {\bibfnamefont {P.}~\bibnamefont {Reshetikhin}}, \bibinfo
  {author} {\bibfnamefont {E.}~\bibnamefont {Druga}}, \bibinfo {author}
  {\bibfnamefont {M.}~\bibnamefont {Bukov}}, \emph {et~al.},\ }\bibfield
  {title} {{\selectlanguage {english}\bibinfo {title} {Critical prethermal
  discrete time crystal created by two-frequency driving}},\ }\href
  {https://doi.org/10.1038/s41567-022-01891-7} {\bibfield  {journal} {\bibinfo
  {journal} {Nature Physics}\ }\textbf {\bibinfo {volume} {19}},\ \bibinfo
  {pages} {407} (\bibinfo {year} {2023})}\BibitemShut {NoStop}%
\bibitem [{\citenamefont {Mäkinen}\ \emph {et~al.}(2023)\citenamefont
  {Mäkinen}, \citenamefont {Autti},\ and\ \citenamefont
  {Eltsov}}]{makinen_magnon_2023}%
  \BibitemOpen
  \bibfield  {author} {\bibinfo {author} {\bibfnamefont {J.~T.}\ \bibnamefont
  {Mäkinen}}, \bibinfo {author} {\bibfnamefont {S.}~\bibnamefont {Autti}},\
  and\ \bibinfo {author} {\bibfnamefont {V.~B.}\ \bibnamefont {Eltsov}},\
  }\href {https://doi.org/10.48550/arXiv.2312.10119} {\bibinfo {title} {Magnon
  {Bose}-{Einstein} condensates: from time crystals and quantum chromodynamics
  to vortex sensing and cosmology}} (\bibinfo {year} {2023})\BibitemShut
  {NoStop}%
\bibitem [{\citenamefont {Euler}\ \emph {et~al.}(2024)\citenamefont {Euler},
  \citenamefont {Braemer}, \citenamefont {Benn},\ and\ \citenamefont
  {Gärttner}}]{euler_metronome_2024}%
  \BibitemOpen
  \bibfield  {author} {\bibinfo {author} {\bibfnamefont {N.}~\bibnamefont
  {Euler}}, \bibinfo {author} {\bibfnamefont {A.}~\bibnamefont {Braemer}},
  \bibinfo {author} {\bibfnamefont {L.}~\bibnamefont {Benn}},\ and\ \bibinfo
  {author} {\bibfnamefont {M.}~\bibnamefont {Gärttner}},\ }\bibfield  {title}
  {\bibinfo {title} {Metronome spin stabilizes time-crystalline dynamics},\
  }\href {https://doi.org/10.1103/PhysRevB.109.224301} {\bibfield  {journal}
  {\bibinfo  {journal} {Phys. Rev. B}\ }\textbf {\bibinfo {volume} {109}},\
  \bibinfo {pages} {224301} (\bibinfo {year} {2024})}\BibitemShut {NoStop}%
\bibitem [{\citenamefont {Zhang}\ \emph {et~al.}(2017)\citenamefont {Zhang},
  \citenamefont {Hess}, \citenamefont {Kyprianidis}, \citenamefont {Becker},
  \citenamefont {Lee}, \citenamefont {Smith}, \citenamefont {Pagano},
  \citenamefont {Potirniche}, \citenamefont {Potter}, \citenamefont
  {Vishwanath} \emph {et~al.}}]{zhangObservationDiscreteTime2017}%
  \BibitemOpen
  \bibfield  {author} {\bibinfo {author} {\bibfnamefont {J.}~\bibnamefont
  {Zhang}}, \bibinfo {author} {\bibfnamefont {P.~W.}\ \bibnamefont {Hess}},
  \bibinfo {author} {\bibfnamefont {A.}~\bibnamefont {Kyprianidis}}, \bibinfo
  {author} {\bibfnamefont {P.}~\bibnamefont {Becker}}, \bibinfo {author}
  {\bibfnamefont {A.}~\bibnamefont {Lee}}, \bibinfo {author} {\bibfnamefont
  {J.}~\bibnamefont {Smith}}, \bibinfo {author} {\bibfnamefont
  {G.}~\bibnamefont {Pagano}}, \bibinfo {author} {\bibfnamefont {I.-D.}\
  \bibnamefont {Potirniche}}, \bibinfo {author} {\bibfnamefont {A.~C.}\
  \bibnamefont {Potter}}, \bibinfo {author} {\bibfnamefont {A.}~\bibnamefont
  {Vishwanath}}, \emph {et~al.},\ }\bibfield  {title} {\bibinfo {title}
  {Observation of a discrete time crystal},\ }\href
  {https://doi.org/10.1038/nature21413} {\bibfield  {journal} {\bibinfo
  {journal} {Nature}\ }\textbf {\bibinfo {volume} {543}},\ \bibinfo {pages}
  {217} (\bibinfo {year} {2017})}\BibitemShut {NoStop}%
\bibitem [{\citenamefont {Choi}\ \emph {et~al.}(2017)\citenamefont {Choi},
  \citenamefont {Choi}, \citenamefont {Landig}, \citenamefont {Kucsko},
  \citenamefont {Zhou}, \citenamefont {Isoya}, \citenamefont {Jelezko},
  \citenamefont {Onoda}, \citenamefont {Sumiya}, \citenamefont {Khemani} \emph
  {et~al.}}]{choiObservationDiscreteTimecrystalline2017}%
  \BibitemOpen
  \bibfield  {author} {\bibinfo {author} {\bibfnamefont {S.}~\bibnamefont
  {Choi}}, \bibinfo {author} {\bibfnamefont {J.}~\bibnamefont {Choi}}, \bibinfo
  {author} {\bibfnamefont {R.}~\bibnamefont {Landig}}, \bibinfo {author}
  {\bibfnamefont {G.}~\bibnamefont {Kucsko}}, \bibinfo {author} {\bibfnamefont
  {H.}~\bibnamefont {Zhou}}, \bibinfo {author} {\bibfnamefont {J.}~\bibnamefont
  {Isoya}}, \bibinfo {author} {\bibfnamefont {F.}~\bibnamefont {Jelezko}},
  \bibinfo {author} {\bibfnamefont {S.}~\bibnamefont {Onoda}}, \bibinfo
  {author} {\bibfnamefont {H.}~\bibnamefont {Sumiya}}, \bibinfo {author}
  {\bibfnamefont {V.}~\bibnamefont {Khemani}}, \emph {et~al.},\ }\bibfield
  {title} {\bibinfo {title} {Observation of discrete time-crystalline order in
  a disordered dipolar many-body system},\ }\href
  {https://doi.org/10.1038/nature21426} {\bibfield  {journal} {\bibinfo
  {journal} {Nature}\ }\textbf {\bibinfo {volume} {543}},\ \bibinfo {pages}
  {221} (\bibinfo {year} {2017})}\BibitemShut {NoStop}%
\bibitem [{\citenamefont {Mi}\ \emph {et~al.}(2022)\citenamefont {Mi},
  \citenamefont {Ippoliti}, \citenamefont {Quintana}, \citenamefont {Greene},
  \citenamefont {Chen}, \citenamefont {Gross}, \citenamefont {Arute},
  \citenamefont {Arya}, \citenamefont {Atalaya}, \citenamefont {Babbush} \emph
  {et~al.}}]{miTimecrystallineEigenstateOrder2022}%
  \BibitemOpen
  \bibfield  {author} {\bibinfo {author} {\bibfnamefont {X.}~\bibnamefont
  {Mi}}, \bibinfo {author} {\bibfnamefont {M.}~\bibnamefont {Ippoliti}},
  \bibinfo {author} {\bibfnamefont {C.}~\bibnamefont {Quintana}}, \bibinfo
  {author} {\bibfnamefont {A.}~\bibnamefont {Greene}}, \bibinfo {author}
  {\bibfnamefont {Z.}~\bibnamefont {Chen}}, \bibinfo {author} {\bibfnamefont
  {J.}~\bibnamefont {Gross}}, \bibinfo {author} {\bibfnamefont
  {F.}~\bibnamefont {Arute}}, \bibinfo {author} {\bibfnamefont
  {K.}~\bibnamefont {Arya}}, \bibinfo {author} {\bibfnamefont {J.}~\bibnamefont
  {Atalaya}}, \bibinfo {author} {\bibfnamefont {R.}~\bibnamefont {Babbush}},
  \emph {et~al.},\ }\bibfield  {title} {\bibinfo {title} {Time-crystalline
  eigenstate order on a quantum processor},\ }\href
  {https://doi.org/10.1038/s41586-021-04257-w} {\bibfield  {journal} {\bibinfo
  {journal} {Nature}\ }\textbf {\bibinfo {volume} {601}},\ \bibinfo {pages}
  {531} (\bibinfo {year} {2022})}\BibitemShut {NoStop}%
\bibitem [{\citenamefont {Frey}\ and\ \citenamefont
  {Rachel}(2022)}]{doi:10.1126/sciadv.abm7652}%
  \BibitemOpen
  \bibfield  {author} {\bibinfo {author} {\bibfnamefont {P.}~\bibnamefont
  {Frey}}\ and\ \bibinfo {author} {\bibfnamefont {S.}~\bibnamefont {Rachel}},\
  }\bibfield  {title} {\bibinfo {title} {Realization of a discrete time crystal
  on 57 qubits of a quantum computer},\ }\bibfield  {journal} {\bibinfo
  {journal} {Science Advances}\ }\textbf {\bibinfo {volume} {8}},\ \href
  {https://doi.org/10.1126/sciadv.abm7652} {10.1126/sciadv.abm7652} (\bibinfo
  {year} {2022})\BibitemShut {NoStop}%
\bibitem [{\citenamefont {Randall}\ \emph {et~al.}(2021)\citenamefont
  {Randall}, \citenamefont {Bradley}, \citenamefont {van~der Gronden},
  \citenamefont {Galicia}, \citenamefont {Abobeih}, \citenamefont {Markham},
  \citenamefont {Twitchen}, \citenamefont {Machado}, \citenamefont {Yao},\ and\
  \citenamefont {Taminiau}}]{doi:10.1126/science.abk0603}%
  \BibitemOpen
  \bibfield  {author} {\bibinfo {author} {\bibfnamefont {J.}~\bibnamefont
  {Randall}}, \bibinfo {author} {\bibfnamefont {C.~E.}\ \bibnamefont
  {Bradley}}, \bibinfo {author} {\bibfnamefont {F.~V.}\ \bibnamefont {van~der
  Gronden}}, \bibinfo {author} {\bibfnamefont {A.}~\bibnamefont {Galicia}},
  \bibinfo {author} {\bibfnamefont {M.~H.}\ \bibnamefont {Abobeih}}, \bibinfo
  {author} {\bibfnamefont {M.}~\bibnamefont {Markham}}, \bibinfo {author}
  {\bibfnamefont {D.~J.}\ \bibnamefont {Twitchen}}, \bibinfo {author}
  {\bibfnamefont {F.}~\bibnamefont {Machado}}, \bibinfo {author} {\bibfnamefont
  {N.~Y.}\ \bibnamefont {Yao}},\ and\ \bibinfo {author} {\bibfnamefont {T.~H.}\
  \bibnamefont {Taminiau}},\ }\bibfield  {title} {\bibinfo {title}
  {Many-body–localized discrete time crystal with a programmable spin-based
  quantum simulator},\ }\href {https://doi.org/10.1126/science.abk0603}
  {\bibfield  {journal} {\bibinfo  {journal} {Science}\ }\textbf {\bibinfo
  {volume} {374}},\ \bibinfo {pages} {1474} (\bibinfo {year}
  {2021})}\BibitemShut {NoStop}%
\bibitem [{\citenamefont {Autti}\ \emph {et~al.}(2018)\citenamefont {Autti},
  \citenamefont {Eltsov},\ and\ \citenamefont
  {Volovik}}]{PhysRevLett.120.215301}%
  \BibitemOpen
  \bibfield  {author} {\bibinfo {author} {\bibfnamefont {S.}~\bibnamefont
  {Autti}}, \bibinfo {author} {\bibfnamefont {V.~B.}\ \bibnamefont {Eltsov}},\
  and\ \bibinfo {author} {\bibfnamefont {G.~E.}\ \bibnamefont {Volovik}},\
  }\bibfield  {title} {\bibinfo {title} {Observation of a time quasicrystal and
  its transition to a superfluid time crystal},\ }\href
  {https://doi.org/10.1103/PhysRevLett.120.215301} {\bibfield  {journal}
  {\bibinfo  {journal} {Phys. Rev. Lett.}\ }\textbf {\bibinfo {volume} {120}},\
  \bibinfo {pages} {215301} (\bibinfo {year} {2018})}\BibitemShut {NoStop}%
\bibitem [{\citenamefont {Autti}\ \emph {et~al.}(2021)\citenamefont {Autti},
  \citenamefont {Heikkinen}, \citenamefont {Mäkinen}, \citenamefont {Volovik},
  \citenamefont {Zavjalov},\ and\ \citenamefont
  {Eltsov}}]{auttiACJosephsonEffect2021}%
  \BibitemOpen
  \bibfield  {author} {\bibinfo {author} {\bibfnamefont {S.}~\bibnamefont
  {Autti}}, \bibinfo {author} {\bibfnamefont {P.~J.}\ \bibnamefont
  {Heikkinen}}, \bibinfo {author} {\bibfnamefont {J.~T.}\ \bibnamefont
  {Mäkinen}}, \bibinfo {author} {\bibfnamefont {G.~E.}\ \bibnamefont
  {Volovik}}, \bibinfo {author} {\bibfnamefont {V.~V.}\ \bibnamefont
  {Zavjalov}},\ and\ \bibinfo {author} {\bibfnamefont {V.~B.}\ \bibnamefont
  {Eltsov}},\ }\bibfield  {title} {\bibinfo {title} {{AC} {Josephson} effect
  between two superfluid time crystals},\ }\href
  {https://doi.org/10.1038/s41563-020-0780-y} {\bibfield  {journal} {\bibinfo
  {journal} {Nature Materials}\ }\textbf {\bibinfo {volume} {20}},\ \bibinfo
  {pages} {171} (\bibinfo {year} {2021})}\BibitemShut {NoStop}%
\bibitem [{\citenamefont {Iemini}\ \emph {et~al.}(2024)\citenamefont {Iemini},
  \citenamefont {Fazio},\ and\ \citenamefont {Sanpera}}]{iemini2023floquet}%
  \BibitemOpen
  \bibfield  {author} {\bibinfo {author} {\bibfnamefont {F.}~\bibnamefont
  {Iemini}}, \bibinfo {author} {\bibfnamefont {R.}~\bibnamefont {Fazio}},\ and\
  \bibinfo {author} {\bibfnamefont {A.}~\bibnamefont {Sanpera}},\ }\bibfield
  {title} {\bibinfo {title} {Floquet time crystals as quantum sensors of ac
  fields},\ }\href {https://doi.org/10.1103/PhysRevA.109.L050203} {\bibfield
  {journal} {\bibinfo  {journal} {Phys. Rev. A}\ }\textbf {\bibinfo {volume}
  {109}},\ \bibinfo {pages} {L050203} (\bibinfo {year} {2024})}\BibitemShut
  {NoStop}%
\bibitem [{\citenamefont {Montenegro}\ \emph {et~al.}(2023)\citenamefont
  {Montenegro}, \citenamefont {Genoni}, \citenamefont {Bayat},\ and\
  \citenamefont {Paris}}]{montenegro_quantum_2023}%
  \BibitemOpen
  \bibfield  {author} {\bibinfo {author} {\bibfnamefont {V.}~\bibnamefont
  {Montenegro}}, \bibinfo {author} {\bibfnamefont {M.~G.}\ \bibnamefont
  {Genoni}}, \bibinfo {author} {\bibfnamefont {A.}~\bibnamefont {Bayat}},\ and\
  \bibinfo {author} {\bibfnamefont {M.~G.~A.}\ \bibnamefont {Paris}},\
  }\bibfield  {title} {{\selectlanguage {english}\bibinfo {title} {Quantum
  metrology with boundary time crystals}},\ }\href
  {https://doi.org/10.1038/s42005-023-01423-6} {\bibfield  {journal} {\bibinfo
  {journal} {Communications Physics}\ }\textbf {\bibinfo {volume} {6}},\
  \bibinfo {pages} {1} (\bibinfo {year} {2023})}\BibitemShut {NoStop}%
\bibitem [{\citenamefont {Gribben}\ \emph {et~al.}(2024)\citenamefont
  {Gribben}, \citenamefont {Sanpera}, \citenamefont {Fazio}, \citenamefont
  {Marino},\ and\ \citenamefont {Iemini}}]{gribben_quantum_2024}%
  \BibitemOpen
  \bibfield  {author} {\bibinfo {author} {\bibfnamefont {D.}~\bibnamefont
  {Gribben}}, \bibinfo {author} {\bibfnamefont {A.}~\bibnamefont {Sanpera}},
  \bibinfo {author} {\bibfnamefont {R.}~\bibnamefont {Fazio}}, \bibinfo
  {author} {\bibfnamefont {J.}~\bibnamefont {Marino}},\ and\ \bibinfo {author}
  {\bibfnamefont {F.}~\bibnamefont {Iemini}},\ }\href@noop {} {\bibinfo {title}
  {{Boundary Time Crystals as AC sensors: enhancements and constraints}}}
  (\bibinfo {year} {2024}),\ \Eprint {https://arxiv.org/abs/2406.06273}
  {arXiv:2406.06273 [quant-ph]} \BibitemShut {NoStop}%
\bibitem [{\citenamefont {Cabot}\ \emph {et~al.}(2024)\citenamefont {Cabot},
  \citenamefont {Carollo},\ and\ \citenamefont
  {Lesanovsky}}]{cabot_continuous_2023}%
  \BibitemOpen
  \bibfield  {author} {\bibinfo {author} {\bibfnamefont {A.}~\bibnamefont
  {Cabot}}, \bibinfo {author} {\bibfnamefont {F.}~\bibnamefont {Carollo}},\
  and\ \bibinfo {author} {\bibfnamefont {I.}~\bibnamefont {Lesanovsky}},\
  }\href {https://doi.org/10.1103/PhysRevLett.132.050801} {\bibinfo {title}
  {Continuous sensing and parameter estimation with the boundary time crystal}}
  (\bibinfo {year} {2024})\BibitemShut {NoStop}%
\bibitem [{\citenamefont {Yousefjani}\ \emph {et~al.}(2024)\citenamefont
  {Yousefjani}, \citenamefont {Sacha},\ and\ \citenamefont
  {Bayat}}]{yousefjani_discrete_2024}%
  \BibitemOpen
  \bibfield  {author} {\bibinfo {author} {\bibfnamefont {R.}~\bibnamefont
  {Yousefjani}}, \bibinfo {author} {\bibfnamefont {K.}~\bibnamefont {Sacha}},\
  and\ \bibinfo {author} {\bibfnamefont {A.}~\bibnamefont {Bayat}},\
  }\href@noop {} {\bibinfo {title} {{Discrete Time Crystal Phase as a Resource
  for Quantum Enhanced Sensing}}} (\bibinfo {year} {2024}),\ \Eprint
  {https://arxiv.org/abs/2405.00328} {arXiv:2405.00328 [quant-ph]} \BibitemShut
  {NoStop}%
\bibitem [{\citenamefont {Moon}\ \emph {et~al.}(2024)\citenamefont {Moon},
  \citenamefont {Schindler}, \citenamefont {Smith}, \citenamefont {Druga},
  \citenamefont {Zhang}, \citenamefont {Bukov},\ and\ \citenamefont
  {Ajoy}}]{moon2024discretetimecrystalsensing}%
  \BibitemOpen
  \bibfield  {author} {\bibinfo {author} {\bibfnamefont {L.~J.~I.}\
  \bibnamefont {Moon}}, \bibinfo {author} {\bibfnamefont {P.~M.}\ \bibnamefont
  {Schindler}}, \bibinfo {author} {\bibfnamefont {R.~J.}\ \bibnamefont
  {Smith}}, \bibinfo {author} {\bibfnamefont {E.}~\bibnamefont {Druga}},
  \bibinfo {author} {\bibfnamefont {Z.-R.}\ \bibnamefont {Zhang}}, \bibinfo
  {author} {\bibfnamefont {M.}~\bibnamefont {Bukov}},\ and\ \bibinfo {author}
  {\bibfnamefont {A.}~\bibnamefont {Ajoy}},\ }\href
  {https://arxiv.org/abs/2410.05625} {\bibinfo {title} {Discrete time crystal
  sensing}} (\bibinfo {year} {2024}),\ \Eprint
  {https://arxiv.org/abs/2410.05625} {arXiv:2410.05625 [quant-ph]} \BibitemShut
  {NoStop}%
\bibitem [{\citenamefont {Shukla}\ \emph {et~al.}(2024)\citenamefont {Shukla},
  \citenamefont {Chotorlishvili}, \citenamefont {Mishra},\ and\ \citenamefont
  {Iemini}}]{shukla2024prethermalfloquettimecrystals}%
  \BibitemOpen
  \bibfield  {author} {\bibinfo {author} {\bibfnamefont {R.~K.}\ \bibnamefont
  {Shukla}}, \bibinfo {author} {\bibfnamefont {L.}~\bibnamefont
  {Chotorlishvili}}, \bibinfo {author} {\bibfnamefont {S.~K.}\ \bibnamefont
  {Mishra}},\ and\ \bibinfo {author} {\bibfnamefont {F.}~\bibnamefont
  {Iemini}},\ }\href {https://arxiv.org/abs/2410.17530} {\bibinfo {title}
  {Prethermal floquet time crystals in chiral multiferroic chains and
  applications as quantum sensors of ac fields}} (\bibinfo {year} {2024}),\
  \Eprint {https://arxiv.org/abs/2410.17530} {arXiv:2410.17530 [quant-ph]}
  \BibitemShut {NoStop}%
\bibitem [{\citenamefont {Estarellas}\ \emph {et~al.}(2020)\citenamefont
  {Estarellas}, \citenamefont {Osada}, \citenamefont {Bastidas}, \citenamefont
  {Renoust}, \citenamefont {Sanaka}, \citenamefont {Munro},\ and\ \citenamefont
  {Nemoto}}]{Estarellas2020}%
  \BibitemOpen
  \bibfield  {author} {\bibinfo {author} {\bibfnamefont {M.~P.}\ \bibnamefont
  {Estarellas}}, \bibinfo {author} {\bibfnamefont {T.}~\bibnamefont {Osada}},
  \bibinfo {author} {\bibfnamefont {V.~M.}\ \bibnamefont {Bastidas}}, \bibinfo
  {author} {\bibfnamefont {B.}~\bibnamefont {Renoust}}, \bibinfo {author}
  {\bibfnamefont {K.}~\bibnamefont {Sanaka}}, \bibinfo {author} {\bibfnamefont
  {W.~J.}\ \bibnamefont {Munro}},\ and\ \bibinfo {author} {\bibfnamefont
  {K.}~\bibnamefont {Nemoto}},\ }\bibfield  {title} {\bibinfo {title}
  {Simulating complex quantum networks with time crystals},\ }\href
  {https://doi.org/10.1126/sciadv.aay8892} {\bibfield  {journal} {\bibinfo
  {journal} {Science Advances}\ }\textbf {\bibinfo {volume} {6}},\ \bibinfo
  {pages} {8892} (\bibinfo {year} {2020})}\BibitemShut {NoStop}%
\bibitem [{\citenamefont {Carollo}\ \emph {et~al.}(2020)\citenamefont
  {Carollo}, \citenamefont {Brandner},\ and\ \citenamefont
  {Lesanovsky}}]{carollo_nonequilibrium_2020}%
  \BibitemOpen
  \bibfield  {author} {\bibinfo {author} {\bibfnamefont {F.}~\bibnamefont
  {Carollo}}, \bibinfo {author} {\bibfnamefont {K.}~\bibnamefont {Brandner}},\
  and\ \bibinfo {author} {\bibfnamefont {I.}~\bibnamefont {Lesanovsky}},\
  }\bibfield  {title} {\bibinfo {title} {Nonequilibrium {Many}-{Body} {Quantum}
  {Engine} {Driven} by {Time}-{Translation} {Symmetry} {Breaking}},\ }\href
  {https://doi.org/10.1103/PhysRevLett.125.240602} {\bibfield  {journal}
  {\bibinfo  {journal} {Phys. Rev. Lett.}\ }\textbf {\bibinfo {volume} {125}},\
  \bibinfo {pages} {240602} (\bibinfo {year} {2020})}\BibitemShut {NoStop}%
\bibitem [{\citenamefont {Bomantara}\ and\ \citenamefont
  {Gong}(2018)}]{bomantara_simulation_2018}%
  \BibitemOpen
  \bibfield  {author} {\bibinfo {author} {\bibfnamefont {R.~W.}\ \bibnamefont
  {Bomantara}}\ and\ \bibinfo {author} {\bibfnamefont {J.}~\bibnamefont
  {Gong}},\ }\bibfield  {title} {\bibinfo {title} {Simulation of
  {Non}-{Abelian} {Braiding} in {Majorana} {Time} {Crystals}},\ }\href
  {https://doi.org/10.1103/PhysRevLett.120.230405} {\bibfield  {journal}
  {\bibinfo  {journal} {Phys. Rev. Lett.}\ }\textbf {\bibinfo {volume} {120}},\
  \bibinfo {pages} {230405} (\bibinfo {year} {2018})}\BibitemShut {NoStop}%
\bibitem [{\citenamefont {Lyu}\ \emph {et~al.}(2020)\citenamefont {Lyu},
  \citenamefont {Choudhury}, \citenamefont {Lv}, \citenamefont {Yan},\ and\
  \citenamefont {Zhou}}]{lyu_eternal_2020}%
  \BibitemOpen
  \bibfield  {author} {\bibinfo {author} {\bibfnamefont {C.}~\bibnamefont
  {Lyu}}, \bibinfo {author} {\bibfnamefont {S.}~\bibnamefont {Choudhury}},
  \bibinfo {author} {\bibfnamefont {C.}~\bibnamefont {Lv}}, \bibinfo {author}
  {\bibfnamefont {Y.}~\bibnamefont {Yan}},\ and\ \bibinfo {author}
  {\bibfnamefont {Q.}~\bibnamefont {Zhou}},\ }\bibfield  {title} {\bibinfo
  {title} {Eternal discrete time crystal beating the {Heisenberg} limit},\
  }\href {https://doi.org/10.1103/PhysRevResearch.2.033070} {\bibfield
  {journal} {\bibinfo  {journal} {Phys. Rev. Res.}\ }\textbf {\bibinfo {volume}
  {2}},\ \bibinfo {pages} {033070} (\bibinfo {year} {2020})}\BibitemShut
  {NoStop}%
\bibitem [{\citenamefont {Bao}\ \emph {et~al.}(2024)\citenamefont {Bao},
  \citenamefont {Xu}, \citenamefont {Song}, \citenamefont {Wang}, \citenamefont
  {Xiang}, \citenamefont {Zhu}, \citenamefont {Chen}, \citenamefont {Jin},
  \citenamefont {Zhu}, \citenamefont {Gao} \emph
  {et~al.}}]{bao_schrodinger_2024}%
  \BibitemOpen
  \bibfield  {author} {\bibinfo {author} {\bibfnamefont {Z.}~\bibnamefont
  {Bao}}, \bibinfo {author} {\bibfnamefont {S.}~\bibnamefont {Xu}}, \bibinfo
  {author} {\bibfnamefont {Z.}~\bibnamefont {Song}}, \bibinfo {author}
  {\bibfnamefont {K.}~\bibnamefont {Wang}}, \bibinfo {author} {\bibfnamefont
  {L.}~\bibnamefont {Xiang}}, \bibinfo {author} {\bibfnamefont
  {Z.}~\bibnamefont {Zhu}}, \bibinfo {author} {\bibfnamefont {J.}~\bibnamefont
  {Chen}}, \bibinfo {author} {\bibfnamefont {F.}~\bibnamefont {Jin}}, \bibinfo
  {author} {\bibfnamefont {X.}~\bibnamefont {Zhu}}, \bibinfo {author}
  {\bibfnamefont {Y.}~\bibnamefont {Gao}}, \emph {et~al.},\ }\bibfield  {title}
  {\bibinfo {title} {{Creating and controlling global
  Greenberger-Horne-Zeilinger entanglement on quantum processors}},\ }\href
  {https://doi.org/10.1038/s41467-024-53140-5} {\bibfield  {journal} {\bibinfo
  {journal} {Nature Commun.}\ }\textbf {\bibinfo {volume} {15}},\ \bibinfo
  {pages} {8823} (\bibinfo {year} {2024})}\BibitemShut {NoStop}%
\bibitem [{\citenamefont {Xu}\ and\ \citenamefont
  {Swingle}(2024)}]{PRXQuantum.5.010201}%
  \BibitemOpen
  \bibfield  {author} {\bibinfo {author} {\bibfnamefont {S.}~\bibnamefont
  {Xu}}\ and\ \bibinfo {author} {\bibfnamefont {B.}~\bibnamefont {Swingle}},\
  }\bibfield  {title} {\bibinfo {title} {Scrambling dynamics and
  out-of-time-ordered correlators in quantum many-body systems},\ }\href
  {https://doi.org/10.1103/PRXQuantum.5.010201} {\bibfield  {journal} {\bibinfo
   {journal} {PRX Quantum}\ }\textbf {\bibinfo {volume} {5}},\ \bibinfo {pages}
  {010201} (\bibinfo {year} {2024})}\BibitemShut {NoStop}%
\bibitem [{\citenamefont {Lewis-Swan}\ \emph {et~al.}(2019)\citenamefont
  {Lewis-Swan}, \citenamefont {Safavi-Naini}, \citenamefont {Kaufman},\ and\
  \citenamefont {Rey}}]{lewis-swanDynamicsQuantumInformation2019}%
  \BibitemOpen
  \bibfield  {author} {\bibinfo {author} {\bibfnamefont {R.~J.}\ \bibnamefont
  {Lewis-Swan}}, \bibinfo {author} {\bibfnamefont {A.}~\bibnamefont
  {Safavi-Naini}}, \bibinfo {author} {\bibfnamefont {A.~M.}\ \bibnamefont
  {Kaufman}},\ and\ \bibinfo {author} {\bibfnamefont {A.~M.}\ \bibnamefont
  {Rey}},\ }\bibfield  {title} {\bibinfo {title} {Dynamics of quantum
  information},\ }\href {https://doi.org/10.1038/s42254-019-0090-y} {\bibfield
  {journal} {\bibinfo  {journal} {Nature Reviews Physics}\ }\textbf {\bibinfo
  {volume} {1}},\ \bibinfo {pages} {627} (\bibinfo {year} {2019})}\BibitemShut
  {NoStop}%
\bibitem [{\citenamefont {Deutsch}(1991)}]{PhysRevA.43.2046}%
  \BibitemOpen
  \bibfield  {author} {\bibinfo {author} {\bibfnamefont {J.~M.}\ \bibnamefont
  {Deutsch}},\ }\bibfield  {title} {\bibinfo {title} {Quantum statistical
  mechanics in a closed system},\ }\href
  {https://doi.org/10.1103/PhysRevA.43.2046} {\bibfield  {journal} {\bibinfo
  {journal} {Phys. Rev. A}\ }\textbf {\bibinfo {volume} {43}},\ \bibinfo
  {pages} {2046} (\bibinfo {year} {1991})}\BibitemShut {NoStop}%
\bibitem [{\citenamefont {Rigol}\ \emph {et~al.}(2008)\citenamefont {Rigol},
  \citenamefont {Dunjko},\ and\ \citenamefont
  {Olshanii}}]{rigolThermalizationItsMechanism2008}%
  \BibitemOpen
  \bibfield  {author} {\bibinfo {author} {\bibfnamefont {M.}~\bibnamefont
  {Rigol}}, \bibinfo {author} {\bibfnamefont {V.}~\bibnamefont {Dunjko}},\ and\
  \bibinfo {author} {\bibfnamefont {M.}~\bibnamefont {Olshanii}},\ }\bibfield
  {title} {\bibinfo {title} {Thermalization and its mechanism for generic
  isolated quantum systems},\ }\href {https://doi.org/10.1038/nature06838}
  {\bibfield  {journal} {\bibinfo  {journal} {Nature}\ }\textbf {\bibinfo
  {volume} {452}},\ \bibinfo {pages} {854} (\bibinfo {year}
  {2008})}\BibitemShut {NoStop}%
\bibitem [{\citenamefont {Srednicki}(1994)}]{PhysRevE.50.888}%
  \BibitemOpen
  \bibfield  {author} {\bibinfo {author} {\bibfnamefont {M.}~\bibnamefont
  {Srednicki}},\ }\bibfield  {title} {\bibinfo {title} {Chaos and quantum
  thermalization},\ }\href {https://doi.org/10.1103/PhysRevE.50.888} {\bibfield
   {journal} {\bibinfo  {journal} {Phys. Rev. E}\ }\textbf {\bibinfo {volume}
  {50}},\ \bibinfo {pages} {888} (\bibinfo {year} {1994})}\BibitemShut
  {NoStop}%
\bibitem [{\citenamefont
  {Swingle}(2018)}]{swingleUnscramblingPhysicsOutoftimeorder2018a}%
  \BibitemOpen
  \bibfield  {author} {\bibinfo {author} {\bibfnamefont {B.}~\bibnamefont
  {Swingle}},\ }\bibfield  {title} {\bibinfo {title} {Unscrambling the physics
  of out-of-time-order correlators},\ }\href
  {https://doi.org/10.1038/s41567-018-0295-5} {\bibfield  {journal} {\bibinfo
  {journal} {Nature Physics}\ }\textbf {\bibinfo {volume} {14}},\ \bibinfo
  {pages} {988} (\bibinfo {year} {2018})}\BibitemShut {NoStop}%
\bibitem [{\citenamefont {Rozenbaum}\ \emph {et~al.}(2017)\citenamefont
  {Rozenbaum}, \citenamefont {Ganeshan},\ and\ \citenamefont
  {Galitski}}]{PhysRevLett.118.086801}%
  \BibitemOpen
  \bibfield  {author} {\bibinfo {author} {\bibfnamefont {E.~B.}\ \bibnamefont
  {Rozenbaum}}, \bibinfo {author} {\bibfnamefont {S.}~\bibnamefont
  {Ganeshan}},\ and\ \bibinfo {author} {\bibfnamefont {V.}~\bibnamefont
  {Galitski}},\ }\bibfield  {title} {\bibinfo {title} {Lyapunov exponent and
  out-of-time-ordered correlator's growth rate in a chaotic system},\ }\href
  {https://doi.org/10.1103/PhysRevLett.118.086801} {\bibfield  {journal}
  {\bibinfo  {journal} {Phys. Rev. Lett.}\ }\textbf {\bibinfo {volume} {118}},\
  \bibinfo {pages} {086801} (\bibinfo {year} {2017})}\BibitemShut {NoStop}%
\bibitem [{\citenamefont {Maldacena}\ \emph {et~al.}(2016)\citenamefont
  {Maldacena}, \citenamefont {Shenker},\ and\ \citenamefont
  {Stanford}}]{maldacenaBoundChaos2016}%
  \BibitemOpen
  \bibfield  {author} {\bibinfo {author} {\bibfnamefont {J.}~\bibnamefont
  {Maldacena}}, \bibinfo {author} {\bibfnamefont {S.~H.}\ \bibnamefont
  {Shenker}},\ and\ \bibinfo {author} {\bibfnamefont {D.}~\bibnamefont
  {Stanford}},\ }\bibfield  {title} {\bibinfo {title} {{A bound on chaos}},\
  }\href {https://doi.org/10.1007/JHEP08(2016)106} {\bibfield  {journal}
  {\bibinfo  {journal} {JHEP}\ }\textbf {\bibinfo {volume} {08}},\ \bibinfo
  {pages} {106}},\ \Eprint {https://arxiv.org/abs/1503.01409} {arXiv:1503.01409
  [hep-th]} \BibitemShut {NoStop}%
\bibitem [{\citenamefont {{Yasuhiro Sekino}}\ and\ \citenamefont {{L.
  Susskind}}(2008)}]{yasuhirosekinoFastScramblers2008a}%
  \BibitemOpen
  \bibfield  {author} {\bibinfo {author} {\bibnamefont {{Yasuhiro Sekino}}}\
  and\ \bibinfo {author} {\bibnamefont {{L. Susskind}}},\ }\bibfield  {title}
  {\bibinfo {title} {Fast scramblers},\ }\href
  {https://doi.org/10.1088/1126-6708/2008/10/065} {\bibfield  {journal}
  {\bibinfo  {journal} {Journal of High Energy Physics}\ }\textbf {\bibinfo
  {volume} {2008}},\ \bibinfo {pages} {065} (\bibinfo {year}
  {2008})}\BibitemShut {NoStop}%
\bibitem [{\citenamefont {Lashkari}\ \emph {et~al.}(2013)\citenamefont
  {Lashkari}, \citenamefont {Stanford}, \citenamefont {Hastings}, \citenamefont
  {Osborne},\ and\ \citenamefont
  {Hayden}}]{lashkariFastScramblingConjecture2013a}%
  \BibitemOpen
  \bibfield  {author} {\bibinfo {author} {\bibfnamefont {N.}~\bibnamefont
  {Lashkari}}, \bibinfo {author} {\bibfnamefont {D.}~\bibnamefont {Stanford}},
  \bibinfo {author} {\bibfnamefont {M.}~\bibnamefont {Hastings}}, \bibinfo
  {author} {\bibfnamefont {T.}~\bibnamefont {Osborne}},\ and\ \bibinfo {author}
  {\bibfnamefont {P.}~\bibnamefont {Hayden}},\ }\bibfield  {title} {\bibinfo
  {title} {Towards the fast scrambling conjecture},\ }\href
  {https://doi.org/10.1007/JHEP04(2013)022} {\bibfield  {journal} {\bibinfo
  {journal} {Journal of High Energy Physics}\ }\textbf {\bibinfo {volume}
  {2013}},\ \bibinfo {pages} {22} (\bibinfo {year} {2013})}\BibitemShut
  {NoStop}%
\bibitem [{\citenamefont {Shenker}\ and\ \citenamefont
  {Stanford}(2014)}]{Shenker:2013pqa}%
  \BibitemOpen
  \bibfield  {author} {\bibinfo {author} {\bibfnamefont {S.~H.}\ \bibnamefont
  {Shenker}}\ and\ \bibinfo {author} {\bibfnamefont {D.}~\bibnamefont
  {Stanford}},\ }\bibfield  {title} {\bibinfo {title} {{Black holes and the
  butterfly effect}},\ }\href {https://doi.org/10.1007/JHEP03(2014)067}
  {\bibfield  {journal} {\bibinfo  {journal} {JHEP}\ }\textbf {\bibinfo
  {volume} {03}},\ \bibinfo {pages} {067}},\ \Eprint
  {https://arxiv.org/abs/1306.0622} {arXiv:1306.0622 [hep-th]} \BibitemShut
  {NoStop}%
\bibitem [{\citenamefont {Swingle}\ \emph {et~al.}(2016)\citenamefont
  {Swingle}, \citenamefont {Bentsen}, \citenamefont {Schleier-Smith},\ and\
  \citenamefont {Hayden}}]{PhysRevA.94.040302}%
  \BibitemOpen
  \bibfield  {author} {\bibinfo {author} {\bibfnamefont {B.}~\bibnamefont
  {Swingle}}, \bibinfo {author} {\bibfnamefont {G.}~\bibnamefont {Bentsen}},
  \bibinfo {author} {\bibfnamefont {M.}~\bibnamefont {Schleier-Smith}},\ and\
  \bibinfo {author} {\bibfnamefont {P.}~\bibnamefont {Hayden}},\ }\bibfield
  {title} {\bibinfo {title} {Measuring the scrambling of quantum information},\
  }\href {https://doi.org/10.1103/PhysRevA.94.040302} {\bibfield  {journal}
  {\bibinfo  {journal} {Phys. Rev. A}\ }\textbf {\bibinfo {volume} {94}},\
  \bibinfo {pages} {040302} (\bibinfo {year} {2016})}\BibitemShut {NoStop}%
\bibitem [{\citenamefont {Mi}\ \emph {et~al.}(2021)\citenamefont {Mi},
  \citenamefont {Roushan}, \citenamefont {Quintana}, \citenamefont {Mandrà},
  \citenamefont {Marshall}, \citenamefont {Neill}, \citenamefont {Arute},
  \citenamefont {Arya}, \citenamefont {Atalaya}, \citenamefont {Babbush} \emph
  {et~al.}}]{doi:10.1126/science.abg5029}%
  \BibitemOpen
  \bibfield  {author} {\bibinfo {author} {\bibfnamefont {X.}~\bibnamefont
  {Mi}}, \bibinfo {author} {\bibfnamefont {P.}~\bibnamefont {Roushan}},
  \bibinfo {author} {\bibfnamefont {C.}~\bibnamefont {Quintana}}, \bibinfo
  {author} {\bibfnamefont {S.}~\bibnamefont {Mandrà}}, \bibinfo {author}
  {\bibfnamefont {J.}~\bibnamefont {Marshall}}, \bibinfo {author}
  {\bibfnamefont {C.}~\bibnamefont {Neill}}, \bibinfo {author} {\bibfnamefont
  {F.}~\bibnamefont {Arute}}, \bibinfo {author} {\bibfnamefont
  {K.}~\bibnamefont {Arya}}, \bibinfo {author} {\bibfnamefont {J.}~\bibnamefont
  {Atalaya}}, \bibinfo {author} {\bibfnamefont {R.}~\bibnamefont {Babbush}},
  \emph {et~al.},\ }\bibfield  {title} {\bibinfo {title} {Information
  scrambling in quantum circuits},\ }\href
  {https://doi.org/10.1126/science.abg5029} {\bibfield  {journal} {\bibinfo
  {journal} {Science}\ }\textbf {\bibinfo {volume} {374}},\ \bibinfo {pages}
  {1479} (\bibinfo {year} {2021})}\BibitemShut {NoStop}%
\bibitem [{\citenamefont {Zhao}\ \emph {et~al.}(2022)\citenamefont {Zhao},
  \citenamefont {Ge}, \citenamefont {Xiang}, \citenamefont {Xue}, \citenamefont
  {Yan}, \citenamefont {Wang}, \citenamefont {Wang}, \citenamefont {Xu},
  \citenamefont {Su}, \citenamefont {Yang} \emph
  {et~al.}}]{PhysRevLett.129.160602}%
  \BibitemOpen
  \bibfield  {author} {\bibinfo {author} {\bibfnamefont {S.~K.}\ \bibnamefont
  {Zhao}}, \bibinfo {author} {\bibfnamefont {Z.-Y.}\ \bibnamefont {Ge}},
  \bibinfo {author} {\bibfnamefont {Z.}~\bibnamefont {Xiang}}, \bibinfo
  {author} {\bibfnamefont {G.~M.}\ \bibnamefont {Xue}}, \bibinfo {author}
  {\bibfnamefont {H.~S.}\ \bibnamefont {Yan}}, \bibinfo {author} {\bibfnamefont
  {Z.~T.}\ \bibnamefont {Wang}}, \bibinfo {author} {\bibfnamefont
  {Z.}~\bibnamefont {Wang}}, \bibinfo {author} {\bibfnamefont {H.~K.}\
  \bibnamefont {Xu}}, \bibinfo {author} {\bibfnamefont {F.~F.}\ \bibnamefont
  {Su}}, \bibinfo {author} {\bibfnamefont {Z.~H.}\ \bibnamefont {Yang}}, \emph
  {et~al.},\ }\bibfield  {title} {\bibinfo {title} {Probing operator spreading
  via floquet engineering in a superconducting circuit},\ }\href
  {https://doi.org/10.1103/PhysRevLett.129.160602} {\bibfield  {journal}
  {\bibinfo  {journal} {Phys. Rev. Lett.}\ }\textbf {\bibinfo {volume} {129}},\
  \bibinfo {pages} {160602} (\bibinfo {year} {2022})}\BibitemShut {NoStop}%
\bibitem [{\citenamefont {Gärttner}\ \emph {et~al.}(2017)\citenamefont
  {Gärttner}, \citenamefont {Bohnet}, \citenamefont {Safavi-Naini},
  \citenamefont {Wall}, \citenamefont {Bollinger},\ and\ \citenamefont
  {Rey}}]{garttner_measuring_2017}%
  \BibitemOpen
  \bibfield  {author} {\bibinfo {author} {\bibfnamefont {M.}~\bibnamefont
  {Gärttner}}, \bibinfo {author} {\bibfnamefont {J.~G.}\ \bibnamefont
  {Bohnet}}, \bibinfo {author} {\bibfnamefont {A.}~\bibnamefont
  {Safavi-Naini}}, \bibinfo {author} {\bibfnamefont {M.~L.}\ \bibnamefont
  {Wall}}, \bibinfo {author} {\bibfnamefont {J.~J.}\ \bibnamefont
  {Bollinger}},\ and\ \bibinfo {author} {\bibfnamefont {A.~M.}\ \bibnamefont
  {Rey}},\ }\bibfield  {title} {\bibinfo {title} {Measuring out-of-time-order
  correlations and multiple quantum spectra in a trapped-ion quantum magnet},\
  }\href {https://doi.org/10.1038/nphys4119} {\bibfield  {journal} {\bibinfo
  {journal} {Nature Physics}\ }\textbf {\bibinfo {volume} {13}},\ \bibinfo
  {pages} {781} (\bibinfo {year} {2017})}\BibitemShut {NoStop}%
\bibitem [{\citenamefont {Braumüller}\ \emph {et~al.}(2022)\citenamefont
  {Braumüller}, \citenamefont {Karamlou}, \citenamefont {Yanay}, \citenamefont
  {Kannan}, \citenamefont {Kim}, \citenamefont {Kjaergaard}, \citenamefont
  {Melville}, \citenamefont {Niedzielski}, \citenamefont {Sung}, \citenamefont
  {Vepsäläinen} \emph {et~al.}}]{braumuller_probing_2022}%
  \BibitemOpen
  \bibfield  {author} {\bibinfo {author} {\bibfnamefont {J.}~\bibnamefont
  {Braumüller}}, \bibinfo {author} {\bibfnamefont {A.~H.}\ \bibnamefont
  {Karamlou}}, \bibinfo {author} {\bibfnamefont {Y.}~\bibnamefont {Yanay}},
  \bibinfo {author} {\bibfnamefont {B.}~\bibnamefont {Kannan}}, \bibinfo
  {author} {\bibfnamefont {D.}~\bibnamefont {Kim}}, \bibinfo {author}
  {\bibfnamefont {M.}~\bibnamefont {Kjaergaard}}, \bibinfo {author}
  {\bibfnamefont {A.}~\bibnamefont {Melville}}, \bibinfo {author}
  {\bibfnamefont {B.~M.}\ \bibnamefont {Niedzielski}}, \bibinfo {author}
  {\bibfnamefont {Y.}~\bibnamefont {Sung}}, \bibinfo {author} {\bibfnamefont
  {A.}~\bibnamefont {Vepsäläinen}}, \emph {et~al.},\ }\bibfield  {title}
  {\bibinfo {title} {Probing quantum information propagation with
  out-of-time-ordered correlators},\ }\href
  {https://doi.org/10.1038/s41567-021-01430-w} {\bibfield  {journal} {\bibinfo
  {journal} {Nature Physics}\ }\textbf {\bibinfo {volume} {18}},\ \bibinfo
  {pages} {172} (\bibinfo {year} {2022})}\BibitemShut {NoStop}%
\bibitem [{\citenamefont {Landsman}\ \emph {et~al.}(2019)\citenamefont
  {Landsman}, \citenamefont {Figgatt}, \citenamefont {Schuster}, \citenamefont
  {Linke}, \citenamefont {Yoshida}, \citenamefont {Yao},\ and\ \citenamefont
  {Monroe}}]{landsman_verified_2019}%
  \BibitemOpen
  \bibfield  {author} {\bibinfo {author} {\bibfnamefont {K.~A.}\ \bibnamefont
  {Landsman}}, \bibinfo {author} {\bibfnamefont {C.}~\bibnamefont {Figgatt}},
  \bibinfo {author} {\bibfnamefont {T.}~\bibnamefont {Schuster}}, \bibinfo
  {author} {\bibfnamefont {N.~M.}\ \bibnamefont {Linke}}, \bibinfo {author}
  {\bibfnamefont {B.}~\bibnamefont {Yoshida}}, \bibinfo {author} {\bibfnamefont
  {N.~Y.}\ \bibnamefont {Yao}},\ and\ \bibinfo {author} {\bibfnamefont
  {C.}~\bibnamefont {Monroe}},\ }\bibfield  {title} {\bibinfo {title} {Verified
  quantum information scrambling},\ }\href
  {https://doi.org/10.1038/s41586-019-0952-6} {\bibfield  {journal} {\bibinfo
  {journal} {Nature}\ }\textbf {\bibinfo {volume} {567}},\ \bibinfo {pages}
  {61} (\bibinfo {year} {2019})}\BibitemShut {NoStop}%
\bibitem [{\citenamefont {Xu}\ and\ \citenamefont
  {Swingle}(2020)}]{xu_accessing_2020}%
  \BibitemOpen
  \bibfield  {author} {\bibinfo {author} {\bibfnamefont {S.}~\bibnamefont
  {Xu}}\ and\ \bibinfo {author} {\bibfnamefont {B.}~\bibnamefont {Swingle}},\
  }\bibfield  {title} {\bibinfo {title} {Accessing scrambling using matrix
  product operators},\ }\href {https://doi.org/10.1038/s41567-019-0712-4}
  {\bibfield  {journal} {\bibinfo  {journal} {Nature Physics}\ }\textbf
  {\bibinfo {volume} {16}},\ \bibinfo {pages} {199} (\bibinfo {year}
  {2020})}\BibitemShut {NoStop}%
\bibitem [{\citenamefont {Luitz}\ and\ \citenamefont
  {Bar~Lev}(2017)}]{PhysRevB.96.020406}%
  \BibitemOpen
  \bibfield  {author} {\bibinfo {author} {\bibfnamefont {D.~J.}\ \bibnamefont
  {Luitz}}\ and\ \bibinfo {author} {\bibfnamefont {Y.}~\bibnamefont
  {Bar~Lev}},\ }\bibfield  {title} {\bibinfo {title} {Information propagation
  in isolated quantum systems},\ }\href
  {https://doi.org/10.1103/PhysRevB.96.020406} {\bibfield  {journal} {\bibinfo
  {journal} {Phys. Rev. B}\ }\textbf {\bibinfo {volume} {96}},\ \bibinfo
  {pages} {020406} (\bibinfo {year} {2017})}\BibitemShut {NoStop}%
\bibitem [{\citenamefont {Bohrdt}\ \emph {et~al.}(2017)\citenamefont {Bohrdt},
  \citenamefont {Mendl}, \citenamefont {Endres},\ and\ \citenamefont
  {Knap}}]{Bohrdt_2017}%
  \BibitemOpen
  \bibfield  {author} {\bibinfo {author} {\bibfnamefont {A.}~\bibnamefont
  {Bohrdt}}, \bibinfo {author} {\bibfnamefont {C.~B.}\ \bibnamefont {Mendl}},
  \bibinfo {author} {\bibfnamefont {M.}~\bibnamefont {Endres}},\ and\ \bibinfo
  {author} {\bibfnamefont {M.}~\bibnamefont {Knap}},\ }\bibfield  {title}
  {\bibinfo {title} {Scrambling and thermalization in a diffusive quantum
  many-body system},\ }\href {https://doi.org/10.1088/1367-2630/aa719b}
  {\bibfield  {journal} {\bibinfo  {journal} {New Journal of Physics}\ }\textbf
  {\bibinfo {volume} {19}},\ \bibinfo {pages} {063001} (\bibinfo {year}
  {2017})}\BibitemShut {NoStop}%
\bibitem [{\citenamefont {Lin}\ and\ \citenamefont
  {Motrunich}(2018)}]{PhysRevB.97.144304}%
  \BibitemOpen
  \bibfield  {author} {\bibinfo {author} {\bibfnamefont {C.-J.}\ \bibnamefont
  {Lin}}\ and\ \bibinfo {author} {\bibfnamefont {O.~I.}\ \bibnamefont
  {Motrunich}},\ }\bibfield  {title} {\bibinfo {title} {Out-of-time-ordered
  correlators in a quantum ising chain},\ }\href
  {https://doi.org/10.1103/PhysRevB.97.144304} {\bibfield  {journal} {\bibinfo
  {journal} {Phys. Rev. B}\ }\textbf {\bibinfo {volume} {97}},\ \bibinfo
  {pages} {144304} (\bibinfo {year} {2018})}\BibitemShut {NoStop}%
\bibitem [{\citenamefont {Burrell}\ and\ \citenamefont
  {Osborne}(2007)}]{PhysRevLett.99.167201}%
  \BibitemOpen
  \bibfield  {author} {\bibinfo {author} {\bibfnamefont {C.~K.}\ \bibnamefont
  {Burrell}}\ and\ \bibinfo {author} {\bibfnamefont {T.~J.}\ \bibnamefont
  {Osborne}},\ }\bibfield  {title} {\bibinfo {title} {Bounds on the speed of
  information propagation in disordered quantum spin chains},\ }\href
  {https://doi.org/10.1103/PhysRevLett.99.167201} {\bibfield  {journal}
  {\bibinfo  {journal} {Phys. Rev. Lett.}\ }\textbf {\bibinfo {volume} {99}},\
  \bibinfo {pages} {167201} (\bibinfo {year} {2007})}\BibitemShut {NoStop}%
\bibitem [{\citenamefont {Swingle}\ and\ \citenamefont
  {Chowdhury}(2017)}]{PhysRevB.95.060201}%
  \BibitemOpen
  \bibfield  {author} {\bibinfo {author} {\bibfnamefont {B.}~\bibnamefont
  {Swingle}}\ and\ \bibinfo {author} {\bibfnamefont {D.}~\bibnamefont
  {Chowdhury}},\ }\bibfield  {title} {\bibinfo {title} {Slow scrambling in
  disordered quantum systems},\ }\href
  {https://doi.org/10.1103/PhysRevB.95.060201} {\bibfield  {journal} {\bibinfo
  {journal} {Phys. Rev. B}\ }\textbf {\bibinfo {volume} {95}},\ \bibinfo
  {pages} {060201} (\bibinfo {year} {2017})}\BibitemShut {NoStop}%
\bibitem [{\citenamefont {Pal}\ and\ \citenamefont
  {Huse}(2010)}]{PhysRevB.82.174411}%
  \BibitemOpen
  \bibfield  {author} {\bibinfo {author} {\bibfnamefont {A.}~\bibnamefont
  {Pal}}\ and\ \bibinfo {author} {\bibfnamefont {D.~A.}\ \bibnamefont {Huse}},\
  }\bibfield  {title} {\bibinfo {title} {Many-body localization phase
  transition},\ }\href {https://doi.org/10.1103/PhysRevB.82.174411} {\bibfield
  {journal} {\bibinfo  {journal} {Phys. Rev. B}\ }\textbf {\bibinfo {volume}
  {82}},\ \bibinfo {pages} {174411} (\bibinfo {year} {2010})}\BibitemShut
  {NoStop}%
\bibitem [{\citenamefont {Abanin}\ \emph {et~al.}(2019)\citenamefont {Abanin},
  \citenamefont {Altman}, \citenamefont {Bloch},\ and\ \citenamefont
  {Serbyn}}]{RevModPhys.91.021001}%
  \BibitemOpen
  \bibfield  {author} {\bibinfo {author} {\bibfnamefont {D.~A.}\ \bibnamefont
  {Abanin}}, \bibinfo {author} {\bibfnamefont {E.}~\bibnamefont {Altman}},
  \bibinfo {author} {\bibfnamefont {I.}~\bibnamefont {Bloch}},\ and\ \bibinfo
  {author} {\bibfnamefont {M.}~\bibnamefont {Serbyn}},\ }\bibfield  {title}
  {\bibinfo {title} {Colloquium: Many-body localization, thermalization, and
  entanglement},\ }\href {https://doi.org/10.1103/RevModPhys.91.021001}
  {\bibfield  {journal} {\bibinfo  {journal} {Rev. Mod. Phys.}\ }\textbf
  {\bibinfo {volume} {91}},\ \bibinfo {pages} {021001} (\bibinfo {year}
  {2019})}\BibitemShut {NoStop}%
\bibitem [{\citenamefont {Nandkishore}\ and\ \citenamefont
  {Huse}(2015)}]{annurev:/content/journals/10.1146/annurev-conmatphys-031214-014726}%
  \BibitemOpen
  \bibfield  {author} {\bibinfo {author} {\bibfnamefont {R.}~\bibnamefont
  {Nandkishore}}\ and\ \bibinfo {author} {\bibfnamefont {D.~A.}\ \bibnamefont
  {Huse}},\ }\bibfield  {title} {\bibinfo {title} {Many-body localization and
  thermalization in quantum statistical mechanics},\ }\href
  {https://doi.org/https://doi.org/10.1146/annurev-conmatphys-031214-014726}
  {\bibfield  {journal} {\bibinfo  {journal} {Annual Review of Condensed Matter
  Physics}\ }\textbf {\bibinfo {volume} {6}},\ \bibinfo {pages} {15} (\bibinfo
  {year} {2015})}\BibitemShut {NoStop}%
\bibitem [{\citenamefont {Vosk}\ \emph {et~al.}(2015)\citenamefont {Vosk},
  \citenamefont {Huse},\ and\ \citenamefont {Altman}}]{PhysRevX.5.031032}%
  \BibitemOpen
  \bibfield  {author} {\bibinfo {author} {\bibfnamefont {R.}~\bibnamefont
  {Vosk}}, \bibinfo {author} {\bibfnamefont {D.~A.}\ \bibnamefont {Huse}},\
  and\ \bibinfo {author} {\bibfnamefont {E.}~\bibnamefont {Altman}},\
  }\bibfield  {title} {\bibinfo {title} {Theory of the many-body localization
  transition in one-dimensional systems},\ }\href
  {https://doi.org/10.1103/PhysRevX.5.031032} {\bibfield  {journal} {\bibinfo
  {journal} {Phys. Rev. X}\ }\textbf {\bibinfo {volume} {5}},\ \bibinfo {pages}
  {031032} (\bibinfo {year} {2015})}\BibitemShut {NoStop}%
\bibitem [{\citenamefont {Altman}(2018)}]{altman_many-body_2018}%
  \BibitemOpen
  \bibfield  {author} {\bibinfo {author} {\bibfnamefont {E.}~\bibnamefont
  {Altman}},\ }\bibfield  {title} {\bibinfo {title} {Many-body localization and
  quantum thermalization},\ }\href {https://doi.org/10.1038/s41567-018-0305-7}
  {\bibfield  {journal} {\bibinfo  {journal} {Nature Physics}\ }\textbf
  {\bibinfo {volume} {14}},\ \bibinfo {pages} {979} (\bibinfo {year}
  {2018})}\BibitemShut {NoStop}%
\bibitem [{\citenamefont {Chen}\ \emph {et~al.}(2017)\citenamefont {Chen},
  \citenamefont {Zhou}, \citenamefont {Huse},\ and\ \citenamefont
  {Fradkin}}]{https://doi.org/10.1002/andp.201600332}%
  \BibitemOpen
  \bibfield  {author} {\bibinfo {author} {\bibfnamefont {X.}~\bibnamefont
  {Chen}}, \bibinfo {author} {\bibfnamefont {T.}~\bibnamefont {Zhou}}, \bibinfo
  {author} {\bibfnamefont {D.~A.}\ \bibnamefont {Huse}},\ and\ \bibinfo
  {author} {\bibfnamefont {E.}~\bibnamefont {Fradkin}},\ }\bibfield  {title}
  {\bibinfo {title} {Out-of-time-order correlations in many-body localized and
  thermal phases},\ }\href
  {https://doi.org/https://doi.org/10.1002/andp.201600332} {\bibfield
  {journal} {\bibinfo  {journal} {Annalen der Physik}\ }\textbf {\bibinfo
  {volume} {529}},\ \bibinfo {pages} {1600332} (\bibinfo {year}
  {2017})}\BibitemShut {NoStop}%
\bibitem [{\citenamefont {Slagle}\ \emph {et~al.}(2017)\citenamefont {Slagle},
  \citenamefont {Bi}, \citenamefont {You},\ and\ \citenamefont
  {Xu}}]{PhysRevB.95.165136}%
  \BibitemOpen
  \bibfield  {author} {\bibinfo {author} {\bibfnamefont {K.}~\bibnamefont
  {Slagle}}, \bibinfo {author} {\bibfnamefont {Z.}~\bibnamefont {Bi}}, \bibinfo
  {author} {\bibfnamefont {Y.-Z.}\ \bibnamefont {You}},\ and\ \bibinfo {author}
  {\bibfnamefont {C.}~\bibnamefont {Xu}},\ }\bibfield  {title} {\bibinfo
  {title} {Out-of-time-order correlation in marginal many-body localized
  systems},\ }\href {https://doi.org/10.1103/PhysRevB.95.165136} {\bibfield
  {journal} {\bibinfo  {journal} {Phys. Rev. B}\ }\textbf {\bibinfo {volume}
  {95}},\ \bibinfo {pages} {165136} (\bibinfo {year} {2017})}\BibitemShut
  {NoStop}%
\bibitem [{\citenamefont {He}\ and\ \citenamefont
  {Lu}(2017)}]{PhysRevB.95.054201}%
  \BibitemOpen
  \bibfield  {author} {\bibinfo {author} {\bibfnamefont {R.-Q.}\ \bibnamefont
  {He}}\ and\ \bibinfo {author} {\bibfnamefont {Z.-Y.}\ \bibnamefont {Lu}},\
  }\bibfield  {title} {\bibinfo {title} {Characterizing many-body localization
  by out-of-time-ordered correlation},\ }\href
  {https://doi.org/10.1103/PhysRevB.95.054201} {\bibfield  {journal} {\bibinfo
  {journal} {Phys. Rev. B}\ }\textbf {\bibinfo {volume} {95}},\ \bibinfo
  {pages} {054201} (\bibinfo {year} {2017})}\BibitemShut {NoStop}%
\bibitem [{\citenamefont {Sahu}\ \emph {et~al.}(2019)\citenamefont {Sahu},
  \citenamefont {Xu},\ and\ \citenamefont {Swingle}}]{PhysRevLett.123.165902}%
  \BibitemOpen
  \bibfield  {author} {\bibinfo {author} {\bibfnamefont {S.}~\bibnamefont
  {Sahu}}, \bibinfo {author} {\bibfnamefont {S.}~\bibnamefont {Xu}},\ and\
  \bibinfo {author} {\bibfnamefont {B.}~\bibnamefont {Swingle}},\ }\bibfield
  {title} {\bibinfo {title} {Scrambling dynamics across a
  thermalization-localization quantum phase transition},\ }\href
  {https://doi.org/10.1103/PhysRevLett.123.165902} {\bibfield  {journal}
  {\bibinfo  {journal} {Phys. Rev. Lett.}\ }\textbf {\bibinfo {volume} {123}},\
  \bibinfo {pages} {165902} (\bibinfo {year} {2019})}\BibitemShut {NoStop}%
\bibitem [{\citenamefont {Bardarson}\ \emph {et~al.}(2012)\citenamefont
  {Bardarson}, \citenamefont {Pollmann},\ and\ \citenamefont
  {Moore}}]{PhysRevLett.109.017202}%
  \BibitemOpen
  \bibfield  {author} {\bibinfo {author} {\bibfnamefont {J.~H.}\ \bibnamefont
  {Bardarson}}, \bibinfo {author} {\bibfnamefont {F.}~\bibnamefont
  {Pollmann}},\ and\ \bibinfo {author} {\bibfnamefont {J.~E.}\ \bibnamefont
  {Moore}},\ }\bibfield  {title} {\bibinfo {title} {Unbounded growth of
  entanglement in models of many-body localization},\ }\href
  {https://doi.org/10.1103/PhysRevLett.109.017202} {\bibfield  {journal}
  {\bibinfo  {journal} {Phys. Rev. Lett.}\ }\textbf {\bibinfo {volume} {109}},\
  \bibinfo {pages} {017202} (\bibinfo {year} {2012})}\BibitemShut {NoStop}%
\bibitem [{\citenamefont {Dumitrescu}\ \emph {et~al.}(2017)\citenamefont
  {Dumitrescu}, \citenamefont {Vasseur},\ and\ \citenamefont
  {Potter}}]{PhysRevLett.119.110604}%
  \BibitemOpen
  \bibfield  {author} {\bibinfo {author} {\bibfnamefont {P.~T.}\ \bibnamefont
  {Dumitrescu}}, \bibinfo {author} {\bibfnamefont {R.}~\bibnamefont
  {Vasseur}},\ and\ \bibinfo {author} {\bibfnamefont {A.~C.}\ \bibnamefont
  {Potter}},\ }\bibfield  {title} {\bibinfo {title} {Scaling theory of
  entanglement at the many-body localization transition},\ }\href
  {https://doi.org/10.1103/PhysRevLett.119.110604} {\bibfield  {journal}
  {\bibinfo  {journal} {Phys. Rev. Lett.}\ }\textbf {\bibinfo {volume} {119}},\
  \bibinfo {pages} {110604} (\bibinfo {year} {2017})}\BibitemShut {NoStop}%
\bibitem [{\citenamefont {Khemani}\ \emph {et~al.}(2017)\citenamefont
  {Khemani}, \citenamefont {Lim}, \citenamefont {Sheng},\ and\ \citenamefont
  {Huse}}]{PhysRevX.7.021013}%
  \BibitemOpen
  \bibfield  {author} {\bibinfo {author} {\bibfnamefont {V.}~\bibnamefont
  {Khemani}}, \bibinfo {author} {\bibfnamefont {S.~P.}\ \bibnamefont {Lim}},
  \bibinfo {author} {\bibfnamefont {D.~N.}\ \bibnamefont {Sheng}},\ and\
  \bibinfo {author} {\bibfnamefont {D.~A.}\ \bibnamefont {Huse}},\ }\bibfield
  {title} {\bibinfo {title} {Critical properties of the many-body localization
  transition},\ }\href {https://doi.org/10.1103/PhysRevX.7.021013} {\bibfield
  {journal} {\bibinfo  {journal} {Phys. Rev. X}\ }\textbf {\bibinfo {volume}
  {7}},\ \bibinfo {pages} {021013} (\bibinfo {year} {2017})}\BibitemShut
  {NoStop}%
\bibitem [{\citenamefont {Sierant}\ \emph {et~al.}(2023)\citenamefont
  {Sierant}, \citenamefont {Lewenstein}, \citenamefont {Scardicchio},\ and\
  \citenamefont {Zakrzewski}}]{PhysRevB.107.115132}%
  \BibitemOpen
  \bibfield  {author} {\bibinfo {author} {\bibfnamefont {P.}~\bibnamefont
  {Sierant}}, \bibinfo {author} {\bibfnamefont {M.}~\bibnamefont {Lewenstein}},
  \bibinfo {author} {\bibfnamefont {A.}~\bibnamefont {Scardicchio}},\ and\
  \bibinfo {author} {\bibfnamefont {J.}~\bibnamefont {Zakrzewski}},\ }\bibfield
   {title} {\bibinfo {title} {Stability of many-body localization in floquet
  systems},\ }\href {https://doi.org/10.1103/PhysRevB.107.115132} {\bibfield
  {journal} {\bibinfo  {journal} {Phys. Rev. B}\ }\textbf {\bibinfo {volume}
  {107}},\ \bibinfo {pages} {115132} (\bibinfo {year} {2023})}\BibitemShut
  {NoStop}%
\bibitem [{\citenamefont {Xu}\ \emph {et~al.}(2021)\citenamefont {Xu},
  \citenamefont {Zhang}, \citenamefont {Han}, \citenamefont {Li}, \citenamefont
  {Xue}, \citenamefont {Liu}, \citenamefont {Jin},\ and\ \citenamefont
  {Yu}}]{xu2021realizingdiscretetimecrystal}%
  \BibitemOpen
  \bibfield  {author} {\bibinfo {author} {\bibfnamefont {H.}~\bibnamefont
  {Xu}}, \bibinfo {author} {\bibfnamefont {J.}~\bibnamefont {Zhang}}, \bibinfo
  {author} {\bibfnamefont {J.}~\bibnamefont {Han}}, \bibinfo {author}
  {\bibfnamefont {Z.}~\bibnamefont {Li}}, \bibinfo {author} {\bibfnamefont
  {G.}~\bibnamefont {Xue}}, \bibinfo {author} {\bibfnamefont {W.}~\bibnamefont
  {Liu}}, \bibinfo {author} {\bibfnamefont {Y.}~\bibnamefont {Jin}},\ and\
  \bibinfo {author} {\bibfnamefont {H.}~\bibnamefont {Yu}},\ }\href
  {https://arxiv.org/abs/2108.00942} {\bibinfo {title} {Realizing discrete time
  crystal in an one-dimensional superconducting qubit chain}} (\bibinfo {year}
  {2021}),\ \Eprint {https://arxiv.org/abs/2108.00942} {arXiv:2108.00942
  [quant-ph]} \BibitemShut {NoStop}%
\bibitem [{\citenamefont {Lee}\ \emph {et~al.}(2019)\citenamefont {Lee},
  \citenamefont {Kim},\ and\ \citenamefont {Kim}}]{PhysRevB.99.184202}%
  \BibitemOpen
  \bibfield  {author} {\bibinfo {author} {\bibfnamefont {J.}~\bibnamefont
  {Lee}}, \bibinfo {author} {\bibfnamefont {D.}~\bibnamefont {Kim}},\ and\
  \bibinfo {author} {\bibfnamefont {D.-H.}\ \bibnamefont {Kim}},\ }\bibfield
  {title} {\bibinfo {title} {Typical growth behavior of the out-of-time-ordered
  commutator in many-body localized systems},\ }\href
  {https://doi.org/10.1103/PhysRevB.99.184202} {\bibfield  {journal} {\bibinfo
  {journal} {Phys. Rev. B}\ }\textbf {\bibinfo {volume} {99}},\ \bibinfo
  {pages} {184202} (\bibinfo {year} {2019})}\BibitemShut {NoStop}%
\bibitem [{\citenamefont {Fan}\ \emph {et~al.}(2017)\citenamefont {Fan},
  \citenamefont {Zhang}, \citenamefont {Shen},\ and\ \citenamefont
  {Zhai}}]{FAN2017707}%
  \BibitemOpen
  \bibfield  {author} {\bibinfo {author} {\bibfnamefont {R.}~\bibnamefont
  {Fan}}, \bibinfo {author} {\bibfnamefont {P.}~\bibnamefont {Zhang}}, \bibinfo
  {author} {\bibfnamefont {H.}~\bibnamefont {Shen}},\ and\ \bibinfo {author}
  {\bibfnamefont {H.}~\bibnamefont {Zhai}},\ }\bibfield  {title} {\bibinfo
  {title} {Out-of-time-order correlation for many-body localization},\ }\href
  {https://doi.org/https://doi.org/10.1016/j.scib.2017.04.011} {\bibfield
  {journal} {\bibinfo  {journal} {Science Bulletin}\ }\textbf {\bibinfo
  {volume} {62}},\ \bibinfo {pages} {707} (\bibinfo {year} {2017})}\BibitemShut
  {NoStop}%
\bibitem [{\citenamefont {Huse}\ \emph {et~al.}(2014)\citenamefont {Huse},
  \citenamefont {Nandkishore},\ and\ \citenamefont
  {Oganesyan}}]{PhysRevB.90.174202}%
  \BibitemOpen
  \bibfield  {author} {\bibinfo {author} {\bibfnamefont {D.~A.}\ \bibnamefont
  {Huse}}, \bibinfo {author} {\bibfnamefont {R.}~\bibnamefont {Nandkishore}},\
  and\ \bibinfo {author} {\bibfnamefont {V.}~\bibnamefont {Oganesyan}},\
  }\bibfield  {title} {\bibinfo {title} {Phenomenology of fully
  many-body-localized systems},\ }\href
  {https://doi.org/10.1103/PhysRevB.90.174202} {\bibfield  {journal} {\bibinfo
  {journal} {Phys. Rev. B}\ }\textbf {\bibinfo {volume} {90}},\ \bibinfo
  {pages} {174202} (\bibinfo {year} {2014})}\BibitemShut {NoStop}%
\bibitem [{\citenamefont {Ros}\ \emph {et~al.}(2015)\citenamefont {Ros},
  \citenamefont {Müller},\ and\ \citenamefont {Scardicchio}}]{ROS2015420}%
  \BibitemOpen
  \bibfield  {author} {\bibinfo {author} {\bibfnamefont {V.}~\bibnamefont
  {Ros}}, \bibinfo {author} {\bibfnamefont {M.}~\bibnamefont {Müller}},\ and\
  \bibinfo {author} {\bibfnamefont {A.}~\bibnamefont {Scardicchio}},\
  }\bibfield  {title} {\bibinfo {title} {Integrals of motion in the many-body
  localized phase},\ }\href
  {https://doi.org/https://doi.org/10.1016/j.nuclphysb.2014.12.014} {\bibfield
  {journal} {\bibinfo  {journal} {Nuclear Physics B}\ }\textbf {\bibinfo
  {volume} {891}},\ \bibinfo {pages} {420} (\bibinfo {year}
  {2015})}\BibitemShut {NoStop}%
\bibitem [{\citenamefont {Tang}\ \emph {et~al.}(2015)\citenamefont {Tang},
  \citenamefont {Iyer},\ and\ \citenamefont {Rigol}}]{PhysRevB.91.161109}%
  \BibitemOpen
  \bibfield  {author} {\bibinfo {author} {\bibfnamefont {B.}~\bibnamefont
  {Tang}}, \bibinfo {author} {\bibfnamefont {D.}~\bibnamefont {Iyer}},\ and\
  \bibinfo {author} {\bibfnamefont {M.}~\bibnamefont {Rigol}},\ }\bibfield
  {title} {\bibinfo {title} {Quantum quenches and many-body localization in the
  thermodynamic limit},\ }\href {https://doi.org/10.1103/PhysRevB.91.161109}
  {\bibfield  {journal} {\bibinfo  {journal} {Phys. Rev. B}\ }\textbf {\bibinfo
  {volume} {91}},\ \bibinfo {pages} {161109} (\bibinfo {year}
  {2015})}\BibitemShut {NoStop}%
\bibitem [{\citenamefont {Serbyn}\ \emph {et~al.}(2013)\citenamefont {Serbyn},
  \citenamefont {Papi\ifmmode~\acute{c}\else \'{c}\fi{}},\ and\ \citenamefont
  {Abanin}}]{PhysRevLett.111.127201}%
  \BibitemOpen
  \bibfield  {author} {\bibinfo {author} {\bibfnamefont {M.}~\bibnamefont
  {Serbyn}}, \bibinfo {author} {\bibfnamefont {Z.}~\bibnamefont
  {Papi\ifmmode~\acute{c}\else \'{c}\fi{}}},\ and\ \bibinfo {author}
  {\bibfnamefont {D.~A.}\ \bibnamefont {Abanin}},\ }\bibfield  {title}
  {\bibinfo {title} {Local conservation laws and the structure of the many-body
  localized states},\ }\href {https://doi.org/10.1103/PhysRevLett.111.127201}
  {\bibfield  {journal} {\bibinfo  {journal} {Phys. Rev. Lett.}\ }\textbf
  {\bibinfo {volume} {111}},\ \bibinfo {pages} {127201} (\bibinfo {year}
  {2013})}\BibitemShut {NoStop}%
\bibitem [{\citenamefont {Imbrie}(2016)}]{imbrie_many-body_2016}%
  \BibitemOpen
  \bibfield  {author} {\bibinfo {author} {\bibfnamefont {J.~Z.}\ \bibnamefont
  {Imbrie}},\ }\bibfield  {title} {\bibinfo {title} {On {Many}-{Body}
  {Localization} for {Quantum} {Spin} {Chains}},\ }\href
  {https://doi.org/10.1007/s10955-016-1508-x} {\bibfield  {journal} {\bibinfo
  {journal} {Journal of Statistical Physics}\ }\textbf {\bibinfo {volume}
  {163}},\ \bibinfo {pages} {998} (\bibinfo {year} {2016})}\BibitemShut
  {NoStop}%
\bibitem [{\citenamefont {Liu}\ \emph {et~al.}(2023{\natexlab{b}})\citenamefont
  {Liu}, \citenamefont {Zhang}, \citenamefont {Hsieh}, \citenamefont {Zhang},\
  and\ \citenamefont {Yao}}]{PhysRevLett.130.120403}%
  \BibitemOpen
  \bibfield  {author} {\bibinfo {author} {\bibfnamefont {S.}~\bibnamefont
  {Liu}}, \bibinfo {author} {\bibfnamefont {S.-X.}\ \bibnamefont {Zhang}},
  \bibinfo {author} {\bibfnamefont {C.-Y.}\ \bibnamefont {Hsieh}}, \bibinfo
  {author} {\bibfnamefont {S.}~\bibnamefont {Zhang}},\ and\ \bibinfo {author}
  {\bibfnamefont {H.}~\bibnamefont {Yao}},\ }\bibfield  {title} {\bibinfo
  {title} {Discrete time crystal enabled by stark many-body localization},\
  }\href {https://doi.org/10.1103/PhysRevLett.130.120403} {\bibfield  {journal}
  {\bibinfo  {journal} {Phys. Rev. Lett.}\ }\textbf {\bibinfo {volume} {130}},\
  \bibinfo {pages} {120403} (\bibinfo {year} {2023}{\natexlab{b}})}\BibitemShut
  {NoStop}%
\bibitem [{\citenamefont {Bull}\ \emph {et~al.}(2022)\citenamefont {Bull},
  \citenamefont {Hallam}, \citenamefont {Papi\ifmmode~\acute{c}\else
  \'{c}\fi{}},\ and\ \citenamefont {Martin}}]{PhysRevLett.129.140602}%
  \BibitemOpen
  \bibfield  {author} {\bibinfo {author} {\bibfnamefont {K.}~\bibnamefont
  {Bull}}, \bibinfo {author} {\bibfnamefont {A.}~\bibnamefont {Hallam}},
  \bibinfo {author} {\bibfnamefont {Z.}~\bibnamefont
  {Papi\ifmmode~\acute{c}\else \'{c}\fi{}}},\ and\ \bibinfo {author}
  {\bibfnamefont {I.}~\bibnamefont {Martin}},\ }\bibfield  {title} {\bibinfo
  {title} {Tuning between continuous time crystals and many-body scars in
  long-range $xyz$ spin chains},\ }\href
  {https://doi.org/10.1103/PhysRevLett.129.140602} {\bibfield  {journal}
  {\bibinfo  {journal} {Phys. Rev. Lett.}\ }\textbf {\bibinfo {volume} {129}},\
  \bibinfo {pages} {140602} (\bibinfo {year} {2022})}\BibitemShut {NoStop}%
\bibitem [{\citenamefont {Maskara}\ \emph
  {et~al.}(2021{\natexlab{b}})\citenamefont {Maskara}, \citenamefont
  {Michailidis}, \citenamefont {Ho}, \citenamefont {Bluvstein}, \citenamefont
  {Choi}, \citenamefont {Lukin},\ and\ \citenamefont
  {Serbyn}}]{PhysRevLett.127.090602}%
  \BibitemOpen
  \bibfield  {author} {\bibinfo {author} {\bibfnamefont {N.}~\bibnamefont
  {Maskara}}, \bibinfo {author} {\bibfnamefont {A.~A.}\ \bibnamefont
  {Michailidis}}, \bibinfo {author} {\bibfnamefont {W.~W.}\ \bibnamefont {Ho}},
  \bibinfo {author} {\bibfnamefont {D.}~\bibnamefont {Bluvstein}}, \bibinfo
  {author} {\bibfnamefont {S.}~\bibnamefont {Choi}}, \bibinfo {author}
  {\bibfnamefont {M.~D.}\ \bibnamefont {Lukin}},\ and\ \bibinfo {author}
  {\bibfnamefont {M.}~\bibnamefont {Serbyn}},\ }\bibfield  {title} {\bibinfo
  {title} {Discrete time-crystalline order enabled by quantum many-body scars:
  Entanglement steering via periodic driving},\ }\href
  {https://doi.org/10.1103/PhysRevLett.127.090602} {\bibfield  {journal}
  {\bibinfo  {journal} {Phys. Rev. Lett.}\ }\textbf {\bibinfo {volume} {127}},\
  \bibinfo {pages} {090602} (\bibinfo {year} {2021}{\natexlab{b}})}\BibitemShut
  {NoStop}%
\bibitem [{\citenamefont {Parker}\ \emph {et~al.}(2019)\citenamefont {Parker},
  \citenamefont {Cao}, \citenamefont {Avdoshkin}, \citenamefont {Scaffidi},\
  and\ \citenamefont {Altman}}]{PhysRevX.9.041017}%
  \BibitemOpen
  \bibfield  {author} {\bibinfo {author} {\bibfnamefont {D.~E.}\ \bibnamefont
  {Parker}}, \bibinfo {author} {\bibfnamefont {X.}~\bibnamefont {Cao}},
  \bibinfo {author} {\bibfnamefont {A.}~\bibnamefont {Avdoshkin}}, \bibinfo
  {author} {\bibfnamefont {T.}~\bibnamefont {Scaffidi}},\ and\ \bibinfo
  {author} {\bibfnamefont {E.}~\bibnamefont {Altman}},\ }\bibfield  {title}
  {\bibinfo {title} {A universal operator growth hypothesis},\ }\href
  {https://doi.org/10.1103/PhysRevX.9.041017} {\bibfield  {journal} {\bibinfo
  {journal} {Phys. Rev. X}\ }\textbf {\bibinfo {volume} {9}},\ \bibinfo {pages}
  {041017} (\bibinfo {year} {2019})}\BibitemShut {NoStop}%
\bibitem [{\citenamefont {Sahu}\ \emph {et~al.}(2024)\citenamefont {Sahu},
  \citenamefont {Bhattacharya},\ and\ \citenamefont {Nath}}]{Sahu:2024urf}%
  \BibitemOpen
  \bibfield  {author} {\bibinfo {author} {\bibfnamefont {H.}~\bibnamefont
  {Sahu}}, \bibinfo {author} {\bibfnamefont {A.}~\bibnamefont {Bhattacharya}},\
  and\ \bibinfo {author} {\bibfnamefont {P.~P.}\ \bibnamefont {Nath}},\
  }\href@noop {} {\bibinfo {title} {{Quantum complexity and localization in
  random quantum circuits}}} (\bibinfo {year} {2024}),\ \Eprint
  {https://arxiv.org/abs/2409.03656} {arXiv:2409.03656 [quant-ph]} \BibitemShut
  {NoStop}%
\bibitem [{\citenamefont {Trigueros}\ and\ \citenamefont
  {Lin}(2022)}]{10.21468/SciPostPhys.13.2.037}%
  \BibitemOpen
  \bibfield  {author} {\bibinfo {author} {\bibfnamefont {F.~B.}\ \bibnamefont
  {Trigueros}}\ and\ \bibinfo {author} {\bibfnamefont {C.-J.}\ \bibnamefont
  {Lin}},\ }\bibfield  {title} {\bibinfo {title} {{Krylov complexity of
  many-body localization: Operator localization in Krylov basis}},\ }\href
  {https://doi.org/10.21468/SciPostPhys.13.2.037} {\bibfield  {journal}
  {\bibinfo  {journal} {SciPost Phys.}\ }\textbf {\bibinfo {volume} {13}},\
  \bibinfo {pages} {037} (\bibinfo {year} {2022})}\BibitemShut {NoStop}%
\bibitem [{\citenamefont {Sahu}(2024)}]{PhysRevA.110.052405}%
  \BibitemOpen
  \bibfield  {author} {\bibinfo {author} {\bibfnamefont {H.}~\bibnamefont
  {Sahu}},\ }\bibfield  {title} {\bibinfo {title} {Information scrambling in
  quantum walks: Discrete-time formulation of krylov complexity},\ }\href
  {https://doi.org/10.1103/PhysRevA.110.052405} {\bibfield  {journal} {\bibinfo
   {journal} {Phys. Rev. A}\ }\textbf {\bibinfo {volume} {110}},\ \bibinfo
  {pages} {052405} (\bibinfo {year} {2024})}\BibitemShut {NoStop}%
\end{thebibliography}%

\end{document}